\documentclass[a4paper,fleqn,usenatbib]{mnras}
\usepackage[T1]{fontenc}
\usepackage{ae,aecompl}
\usepackage{graphicx}	% Including figure files
\usepackage{amsmath}	% Advanced maths commands
\usepackage{amssymb}	% Extra maths symbols
\usepackage{makecell}
\usepackage{multirow}
\newcommand{\kms}{km s$^{-1}$}
\title[De-projection of axi-symmetric expanding circumstellar envelopes]{De-projection of radio observations of axi-symmetric expanding circumstellar envelopes}
\author[P.T. Nhung et al.]{{P.T. Nhung$^1$\thanks{E-mail: pttnhung@vnsc.org.vn}, D.T. Hoai$^1$, P. Tuan-Anh$^1$, P. Darriulat$^1$, T. Le Bertre$^2$,}
\newauthor{J.M. Winters$^3$, P.N. Diep$^1$ and N.T. Phuong$^1$}
\\
$^1$Department of Astrophysics, Vietnam National Space Center (VNSC), Vietnam Academy of Science and Technology (VAST), \\18 Hoang Quoc Viet, Ha Noi, Viet Nam\\
$^2$LERMA, UMR 8112, CNRS and Observatoire de Paris, PSL Research University, 61 av. de l'Observatoire, F-75014 Paris, France\\
$^3$IRAM, 300 rue de la Piscine, Domaine Universitaire, F-38406 St. Martin d'H\'{e}res, France\\
}
\date{Accepted XXX. Received YYY; in original form ZZZ}

\pubyear{2018}
\begin{document}
\label{firstpage}
\pagerange{\pageref{firstpage}--\pageref{lastpage}}
\maketitle

\begin{abstract}
  The problem of de-projection of radio line observations of axi-symmetric expanding circumstellar
  envelopes is studied with the aim of easing their analysis in terms of physics models.
  The arguments developed rest on the remark that, in principle, when the wind velocity
  distribution is known, the effective emissivity can be calculated at any point in space.
  The paper provides a detailed study of how much this is true in practice. The wind velocity
  distribution assumed to be axi-symmetric and in expansion, is described by four parameters:
  the angles defining the orientation of the symmetry axis, an overall velocity scale and
  a parameter measuring the elongation (prolateness) of the distribution. Tools are developed
  that allow for measuring, or at least constraining, each of the four parameters. The use
  of effective emissivity as relevant quantity, rather than temperature and density being
  considered separately, implies important assumptions and simplifications meaning that the
  approach being considered here is only a preliminary to, and by no means a replacement
  for, a physics analysis accounting for radiative transfer and hydrodynamics arguments.
  While most considerations are developed using simulated observations as examples, two
  case studies (EP Aqr, observed with ALMA, and RS Cnc, with NOEMA) are presented that
  illustrate their usefulness in practical cases.
\end{abstract}


%%%%%%%%%%%%%%%%%%%%%%%%%%%%%%%%%%%%%%%%%%%%%%%%%%

%%%%%%%%%%%%%%%%% BODY OF PAPER %%%%%%%%%%%%%%%%%%

\section{INTRODUCTION}

%This is a simple template for authors to write new MNRAS papers.

Recent years have seen high quality observations of molecular line emissions from evolved
stars become available. Such observations, in particular from NOEMA (NOrthern Extended Millimeter Array)
and from the Atacama Large Millimeter/submillimeter Array (ALMA), offer a spatial and spectral resolution calling
for analysis methods making the best possible use of it. The present work addresses the
case of expanding circumstellar envelopes of evolved stars, particularly stars populating
the Asymptotic Giant Branch \citep[AGB, for a recent review see][and references therein]{Hofner2018}.
Such envelopes have shapes that often evolve from spherical to
axi-symmetric morphology, providing the seed for possibly more irregular configurations
later observed in post-AGB stars and Planetary Nebulae. The physics governing the breaking
of spherical symmetry is currently the subject of active research. In the case of binaries,
attraction from the companion has been shown to play an important role. Yet, many unanswered
questions, such as the role played by magnetic fields, remain to be elucidated. \\

The analysis of radio observations of molecular line emissions requires, as a preliminary,
a de-projection in space. This is a largely under-determined problem: only two out of three
position coordinates are measured, those in the sky plane; the position along the line of
sight is unknown. And only one out of three velocity components is measured, that along
the line of sight, from the observed Doppler shift.  In a recent paper \citep[]{Diep2016} 
we made general considerations on the problem of de-projection, with particular emphasis on
the differentiation between expansion and rotation and on the use of Position-Velocity (PV)
diagrams. We were then addressing issues related to both the physics of proto-stars and of
evolved stars. In the present work, we concentrate instead on expanding circumstellar envelopes.
The aim is to shed new light on the problem of de-projection in as simple terms as possible
and to understand in depth the difficulties that its solution needs to face. To do so, we
deliberately ignore complications such as arising from optical thickness or from the possible
presence of rotation, and more generally from any form of departure from exact axi-symmetry
and exact radial expansion. We mostly exploit the simplification offered by the constraint of
axi-symmetry (one relation) and from the hypothesis of radial expansion (two relations),
helping with the solution of the problem of de-projection.\\

The considerations that follow have no ambition at replacing the physics analysis required
by an in-depth understanding of the physics mechanisms at play. They are simply meant as a
useful preliminary step, providing helpful tools and possibly inspiring considerations on the issue of de-projection.  
\section{THE FRAMEWORK}\label{sec.2}
%%%%%%%%%%%%  star from here  %%%%%%%%%%%%%%%%%%

\begin{figure}
\centering
\includegraphics[width=0.5\textwidth,trim=0.cm 0.cm 0.cm 0.cm,clip]{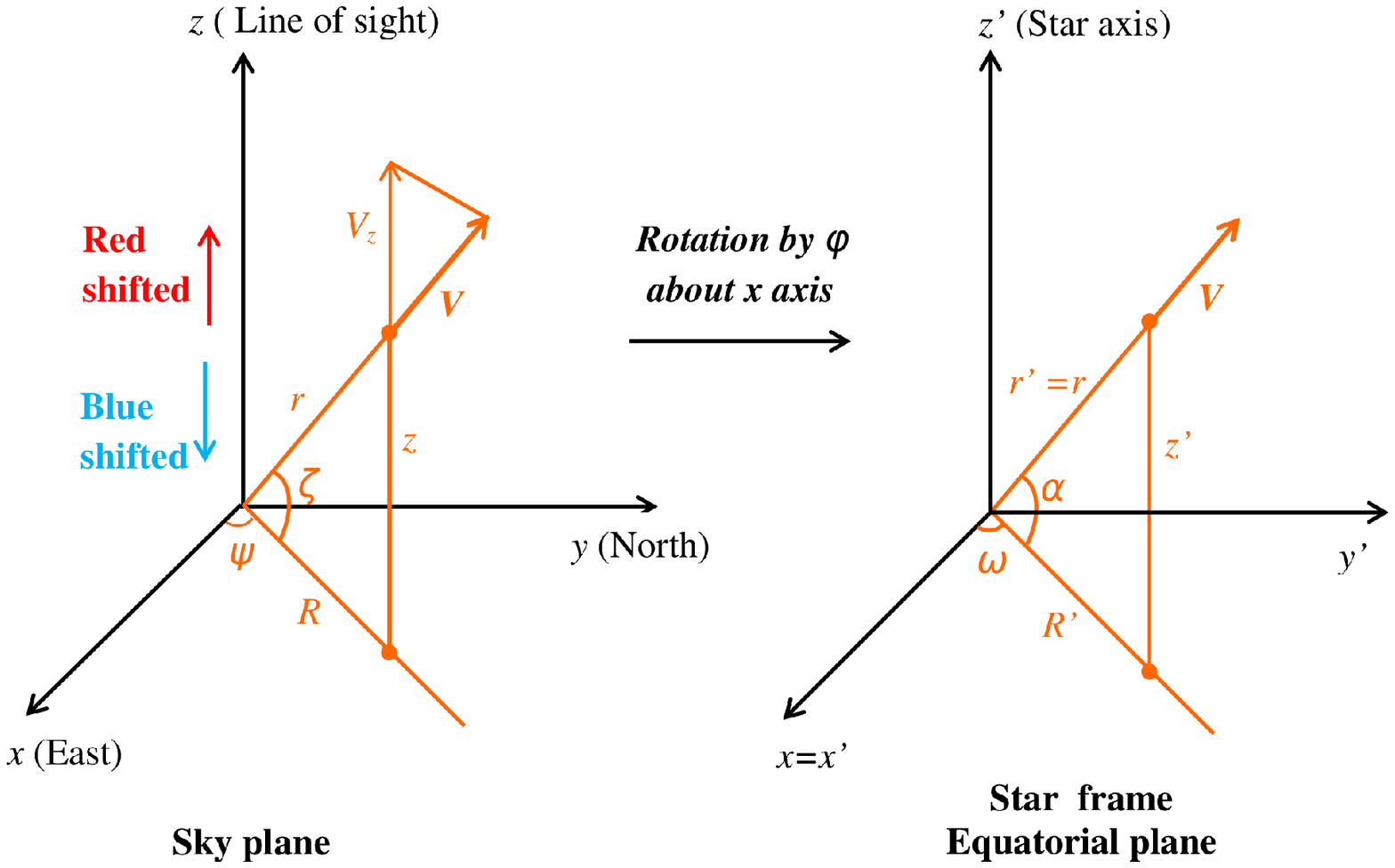}
\caption{Coordinate systems for $\theta=0$. Relations \ref{eq.5} transform from the sky frame (left)
  to the star frame (right).}
\label{fig1}
\end{figure}

We use coordinates ($x,y,z$) attached to the sky plane and ($x',y',z'$)  attached to the star (Figure \ref{fig1}).
The $z$ axis is parallel to the line of sight, pointing away from Earth, while $x$ is pointing
east and $y$ north. The $z'$ axis is the symmetry axis of the star morphology and kinematics,
making an angle $\varphi$ with the line of sight. Its projection on the sky plane makes an angle
$\theta$ (position angle) with the $y$ axis, where $\theta$ is the angle between the $x$ axis and the $x'$
axis, the latter taken to be in the ($x,y$) plane. The transformation relations between the two systems
of coordinates read
\begin{equation}
  \begin{split}
    x'&= x\cos\theta+y\sin\theta \\
    y'&=(-x\sin\theta + y\cos\theta) \cos\varphi + z\sin\varphi \\
    z'&= -(- x\sin\theta + y\cos\theta ) \sin\varphi + z\cos\varphi\\    
    x&=x'\cos\theta-(y'\cos\varphi-z'\sin\varphi)\sin\theta\\
    y&=(y'\cos\varphi - z' \sin\varphi) \cos\theta + x' \sin\theta\\
    z&=y'\sin\varphi + z'\cos\varphi
  \end{split}
\end{equation}

\begin{figure*}
\centering
\includegraphics[height=7cm,trim=0.cm 0.cm 0.cm 0.cm,clip]{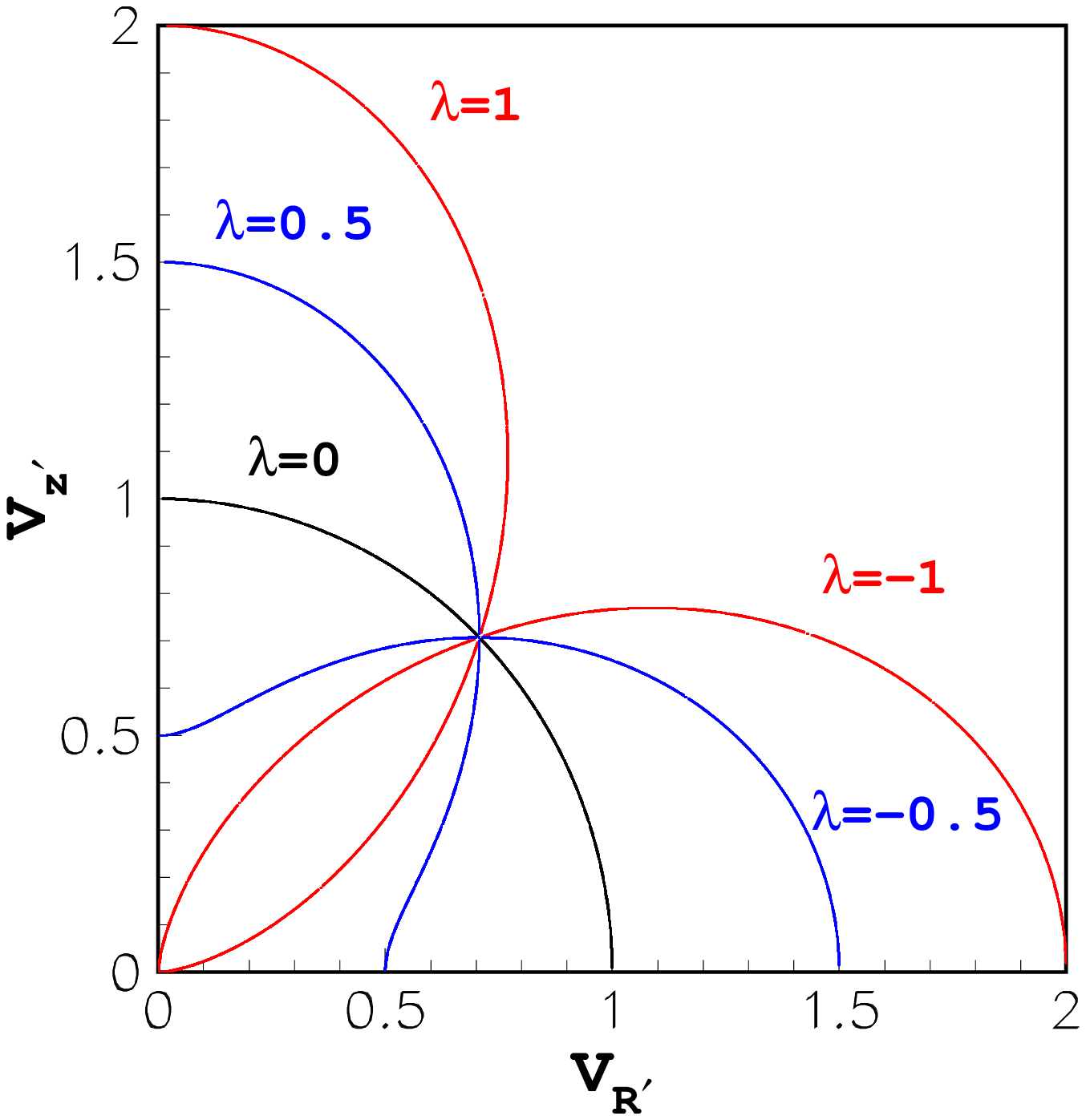}
\includegraphics[height=6.7cm,trim=2.cm 0.5cm 0.cm 0.cm,clip]{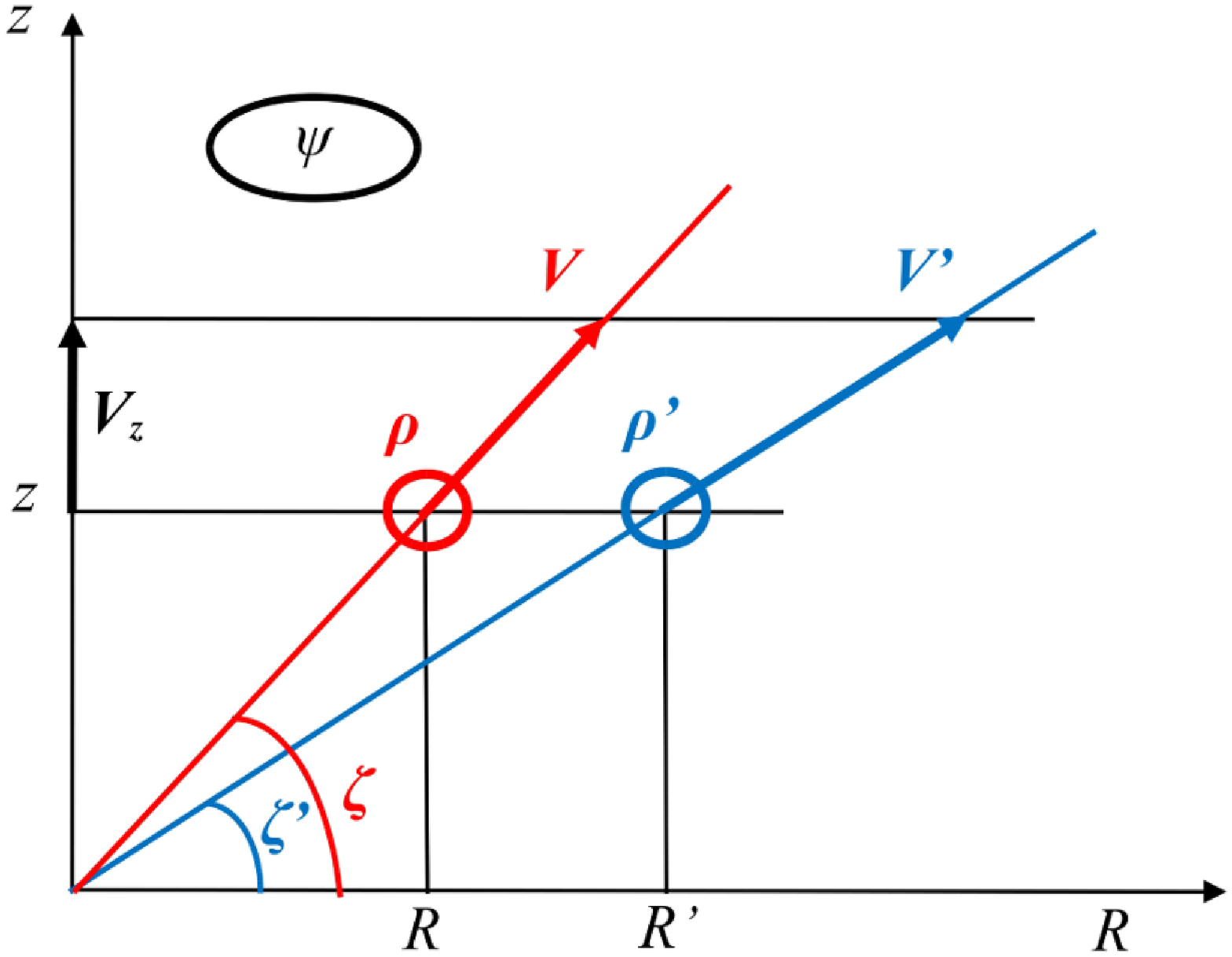}
\caption{Left: polar diagrams of $V$ in the upper meridian quadrant ($R'$ in abscissa and $z'$ in ordinate)
  for different values of the prolateness parameter $\lambda$. Right: Deprojection in the ($R,z$) plane (see text).}
\label{fig2}
\end{figure*}

However, in much of what follows, we redefine the $y$ axis as the projection of the star axis
on the sky plane, which is equivalent to setting $\theta=0$. In this case $x=x'$ and the
transformation relations between ($y,z$) and ($y',z'$) read:
\begin{align}\label{eq.2}
	\begin{split}
		y'=y\cos\varphi + z\sin\varphi &\qquad 		y=y'\cos\varphi - z' \sin\varphi \\
		 z'=-y\sin\varphi + z\cos\varphi &\qquad	z=y' \sin\varphi + z' \cos\varphi
	\end{split}
\end{align}
In order to illustrate our arguments as simply as possible we use a model (in the remainder
of the article we refer to it as \textquotedblleft the simple model\textquotedblright) in
which the wind velocity is radial and independent of the distance from the star with a
dependence on star latitude $\alpha$ of the form
\begin{equation}\label{eq.3}
	V=V_{pole}\sin^2\alpha + V_{eq} \cos^2\alpha=V_0( 1 - \lambda \cos2\alpha)
\end{equation}
with the polar velocity $V_{pole}=V_0(1+\lambda)$ and the equatorial velocity $V_{eq}=V_0(1 - \lambda)$
(Figure \ref{fig2} left). For $\lambda$ between 0 and 1 one obtains prolate velocity distributions (bipolar
outflow) and for $\lambda$ between $-1$ and $0$ oblate velocity distributions (equatorial outflow),
$\lambda$=0 corresponding to an isotropic (spherical) velocity distribution. Here $V_0$ defines the
velocity scale but its precise value is irrelevant; in practice, we set it at 5 km s$^{-1}$ for the
purpose of illustration. The Doppler velocity at space point ($x,y,z$) is
\begin{equation}\label{eq.4}
	V_{z}=V \sin\zeta
\end{equation}
where $\zeta$   is the angle between the ($x,y,z$) direction and the plane of the sky.
We also define $R=\sqrt{x^2+y^2}=z/{\tan\zeta}$ and $r=\sqrt{R^2+z^2}=z/{\sin\zeta}=R/{\cos\zeta}.$\\
From Relations \ref{eq.2}, \ref{eq.3} and \ref{eq.4}, using  $\sin\alpha=z'/r$ and replacing
$z'$ by its expression in Relation \ref{eq.2} we can express the Doppler velocity as a function
of $\zeta$, $\varphi$ and $\psi$, defined as the position angle measured with respect to the projection
of the star axis on the sky plane, $y=R\cos\psi$:
\begin{equation}\label{eq.5}
  \begin{split}
    V_z &= V_0 (z/r)[1 - \lambda (1 - 2\sin^2\alpha)]\\
    &=V_0(z/r)[1 - \lambda + 2 \lambda(z \cos\varphi  - y \sin\varphi)^2/r^2] \\
  \end{split}
\end{equation}
\begin{align}
  V_z  = &V_0 \sin\zeta [ 1 - \lambda + 2 \lambda ( \sin\zeta \cos\varphi - \cos\zeta \cos\psi \sin\varphi)^2]\nonumber\\
  = &V_0 [( 1 - \lambda ) \sin\zeta + 2 \lambda \sin^3\zeta \cos^2\varphi\nonumber\\
  &- 4 \lambda \sin^2\zeta \cos\zeta \sin\varphi \cos\varphi \cos\psi \nonumber\\
  & + 2\lambda \sin\zeta \cos^2\zeta \sin^2\varphi \cos^2\psi] \tag{5a}\nonumber
\end{align}

Taking the derivative along the line of sight ($R$ and $\psi$ being fixed) and using the identities
$d/dz=[d/d(\sin\zeta)][d(\sin\zeta)/dz]=r^{-1} (\cos^2 \zeta) d/d(\sin\zeta)$ and $ d(\cos\zeta)/d(\sin\zeta)= - \tan\zeta$ we obtain:
\begin{align*}
	rdV_z/dz=V_0[(1 - \lambda)\cos^2\zeta + 2\lambda P]  \tag{5b}
\end{align*}
with
\begin{align*}
  P=&3\cos^2\varphi \sin^2\zeta \cos^2\zeta\\
  &-\cos\psi \sin2\varphi \cos\zeta \sin\zeta(2\cos^2\zeta-\sin^2\zeta) \\
	&+ \cos^2\psi \sin^2\varphi \cos^2\zeta(\cos^2\zeta-2\sin^2\zeta) \tag{5c}
\end{align*}

Radio astronomy measurements are in the form of a data-cube with elements $f(x,y,V_z$) measuring the brightness
along the line ($x,y$) normal to the sky plane at Doppler velocity $V_z$. It is convenient to define
the effective emissivity at space point ($x,y,z$) as 
\begin{equation}\label{eq.6}
	\rho(x,y,z)=f(x,y,V_z)dV_z/dz	
\end{equation}
and the measured intensity as 
\begin{equation}\label{eq.7}
	F(x,y)= \int{f(x,y,V_z)dV_z}=\int{\rho(x,y,z)dz}
\end{equation}

As mentioned in the introduction, the assumption of pure radial velocity and of axi-symmetry should help
with the de-projection of the effective emissivity using Relation~\ref{eq.6}. How much this is true in
practice is the subject of the present article.\\

\begin{figure*}
\centering\includegraphics[height=6cm,trim=0.cm .5cm 1.cm 1.cm,clip]{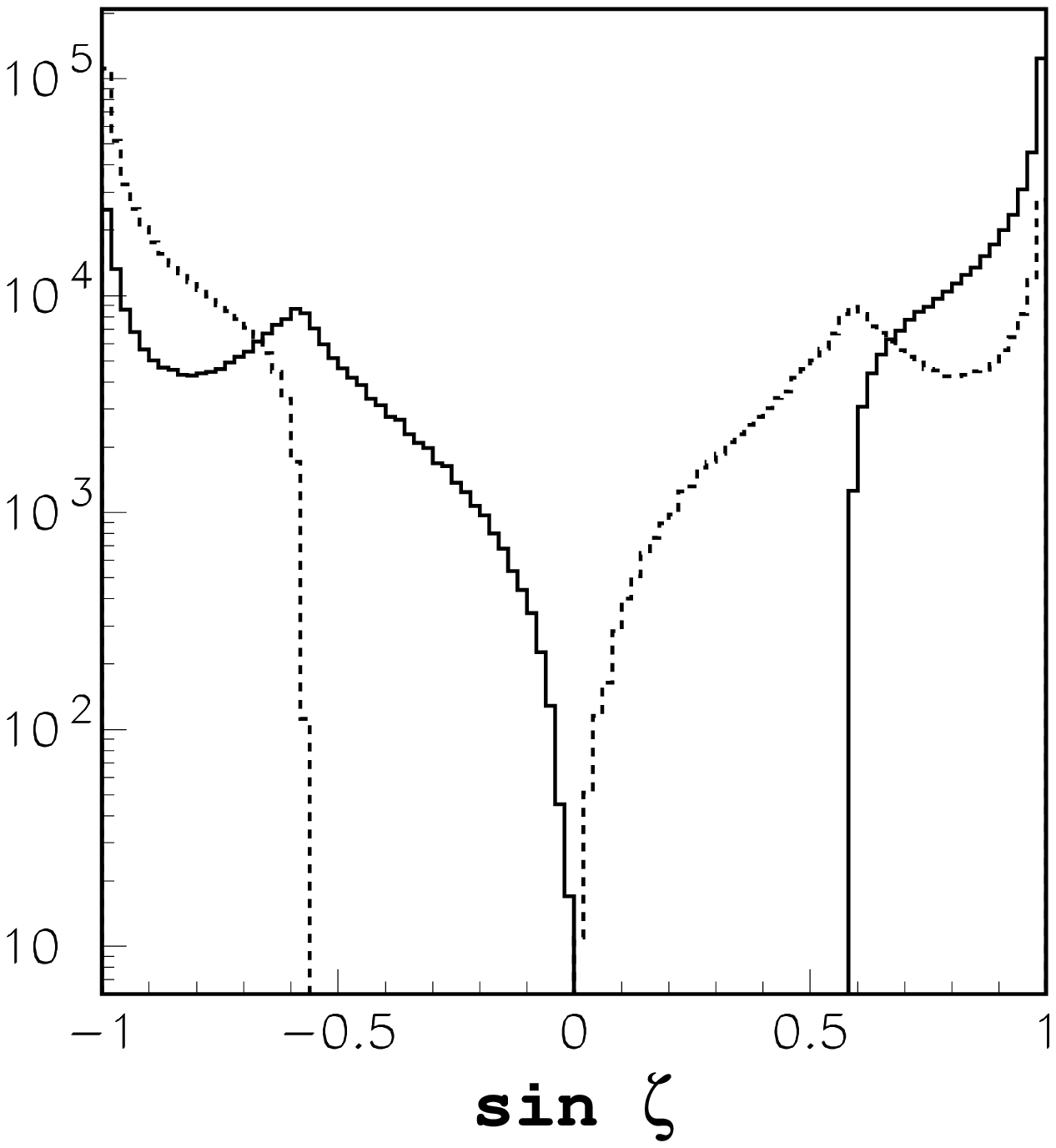}
\includegraphics[height=6cm,trim=.8cm .5cm 0.cm 1.cm,clip]{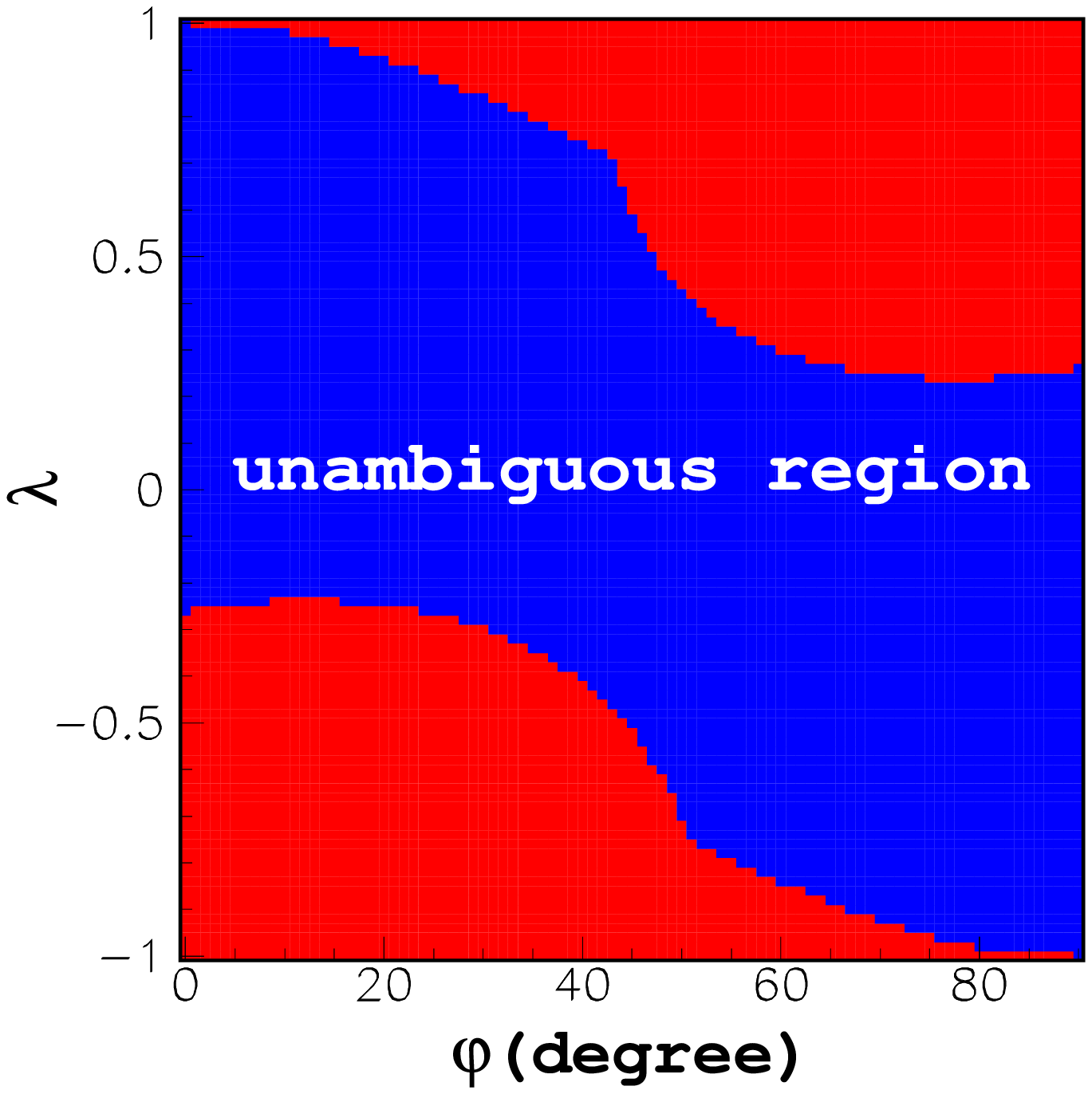}
\caption{Left: $\sin\zeta$ (abscissa) distribution of the extrema of the $V_z$ for a million of uniformly distributed $(\lambda,\varphi,\psi)$ triplets. Dashed line is for $\lambda$<0 (oblate) and solid line for $\lambda$>0 (prolate). Right: region (in blue) of the $\lambda$ (ordinate) vs $\varphi$ (abscissa) plane where de-projection is unambiguous, $z$ increasing monotonically with $V_z$ in the region $|\sin\zeta|<0.9$.  }
\label{fig3}
\end{figure*}

A number of symmetries are apparent in Relations \ref{eq.5}. Changing $\varphi$ in $-\varphi$  and $\psi$ in
$180^\circ - \psi$ leaves $V_z$ invariant, so does also changing $\psi$ in $-\psi$; changing
$\varphi$ in $180^\circ - \varphi$ and $\zeta$ in $-\zeta$ changes $V_z$ in $- V_z$. Accordingly, it is
sufficient to limit the ranges of $\varphi$ and $\psi$ to respectively [$0^\circ, 90^\circ$] and [$0^\circ, 180^\circ$]. \\

\begin{figure*}
\centering
\includegraphics[width=0.8\textwidth,trim=0.cm 0.cm 0.cm 0.cm,clip]{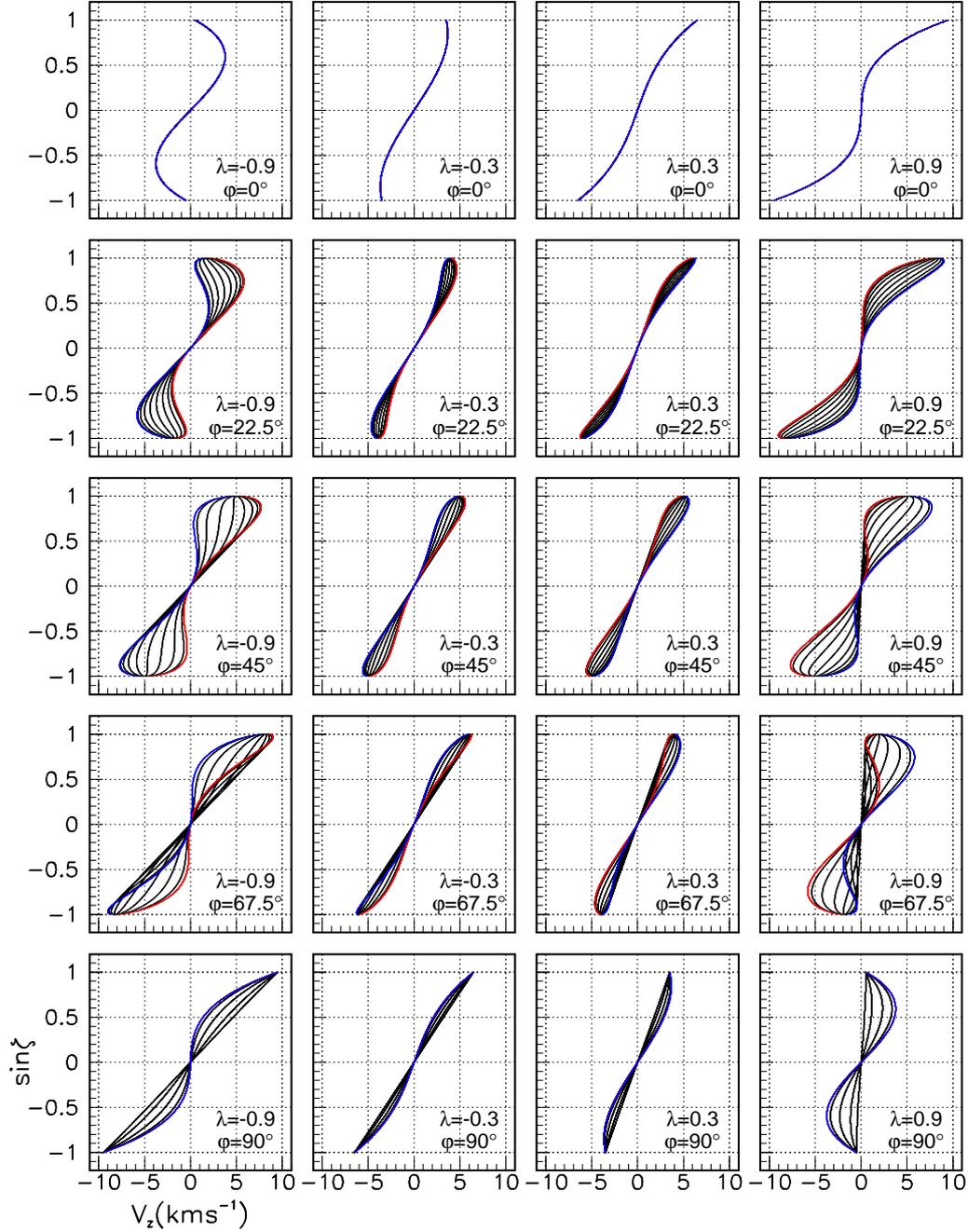}
\caption{Dependence of $\sin\zeta$ (ordinate) on $V_z$ (abscissa) for different values of $\psi$ (from 0$^\circ$ to 180$^\circ$ in steps of 20$^\circ$). Panels are in five rows of $\varphi$ (up down from 0$^\circ$ to 90$^\circ$ in steps of 22.5$^\circ$) and in columns of $\lambda$ (left to right from $-0.9$ to $+0.9$ in steps of 0.6).}
\label{fig4}
\end{figure*}
A remarkable consequence of the above relations is the ability to estimate the $r$-dependence of the
effective emissivity independently from the form chosen for the wind velocity ($V_0,\lambda,\varphi$)
as long as $V$ does not depend on $r$ (Figure~\ref{fig2} right). Indeed, consider a line passing by the star
and having direction ($\psi,\zeta$), namely making an angle $\zeta$ with the plane of the sky and
projecting on it at position angle $\psi$; this line corresponds to a single value of $V_z$, independent
of both $R$ and $r$ as long as the radial expansion velocity is constant on it, namely independent of $r$.
Therefore, as long as both the real wind velocity, $V(\psi,\zeta)$, and the wind velocity used for
de-projection, $V'(\psi,\zeta)$ are both independent of $r$ (but depend on $\psi$ and $\zeta$, generally
in different ways)  the de-projected line is also a straight line, having the same projection on the
plane of the sky; its direction ($\psi',\zeta'$) is related to the direction ($\psi,\zeta$)  by the
relations $\psi'=\psi$ and $V'(\psi,\zeta')\sin\zeta'=V(\psi,\zeta)\sin\zeta$.  De-projection simply
transforms $\zeta$ into $\zeta'$ and the ratio $z'/z$ is independent of $R$. As a result, the true
effective emissivity $\rho$ and the de-projected effective emissivity $\rho'$ have the same dependence
on $R$ on each of these two lines, ($\psi,\zeta$) for the former and ($\psi,\zeta'$) for the latter:
they are proportional to a same data-cube element $f(x,y,V_z)$, independently from $R$ . To the extent
that the true wind velocity and the wind velocity used for de-projection are not too different, the
$r$-dependence of the de-projected effective emissivity is similar to the $r$-dependence of the true
effective emissivity, both being dominated by the $R$ dependence of the Doppler velocity spectrum.
\section{GENERAL CONSIDERATIONS}\label{sec.3}
De-projection implies using Relation \ref{eq.6} to calculate the effective emissivity $\rho$($x,y,z$) from the measured brightness $f(x,y,V_z)$ by associating to each value $V_z$ of the measured Doppler velocity spectrum a point ($x,y,z$) in space. This is only possible if the relation giving $V_z$ as a function of $z$ can be inverted into a relation giving $z$ as a function of $V_z$. In general, an extremum of the dependence of $V_z$ on $z$ will generate in its vicinity two values of $z$ for a same value of $V_z$. In principle, this should prevent de-projection as one does not know how to share the brightness measured at Doppler velocity $V_z$ between the two corresponding space points; this issue will be discussed in some detail in Section \ref{sec.6}. As the relation between $z$ and $\sin\zeta$ is one-to-one, $z=R\tan\zeta=R\sin\zeta / \sqrt{1-\sin^2\zeta}$, the extrema of $V_z$ vs $z$ are the same as of $V_z$ vs $\sin\zeta$.\\

\begin{figure*}
\centering
\includegraphics[width=0.4\textwidth,trim=0.cm 0cm 0.cm 0.cm,clip]{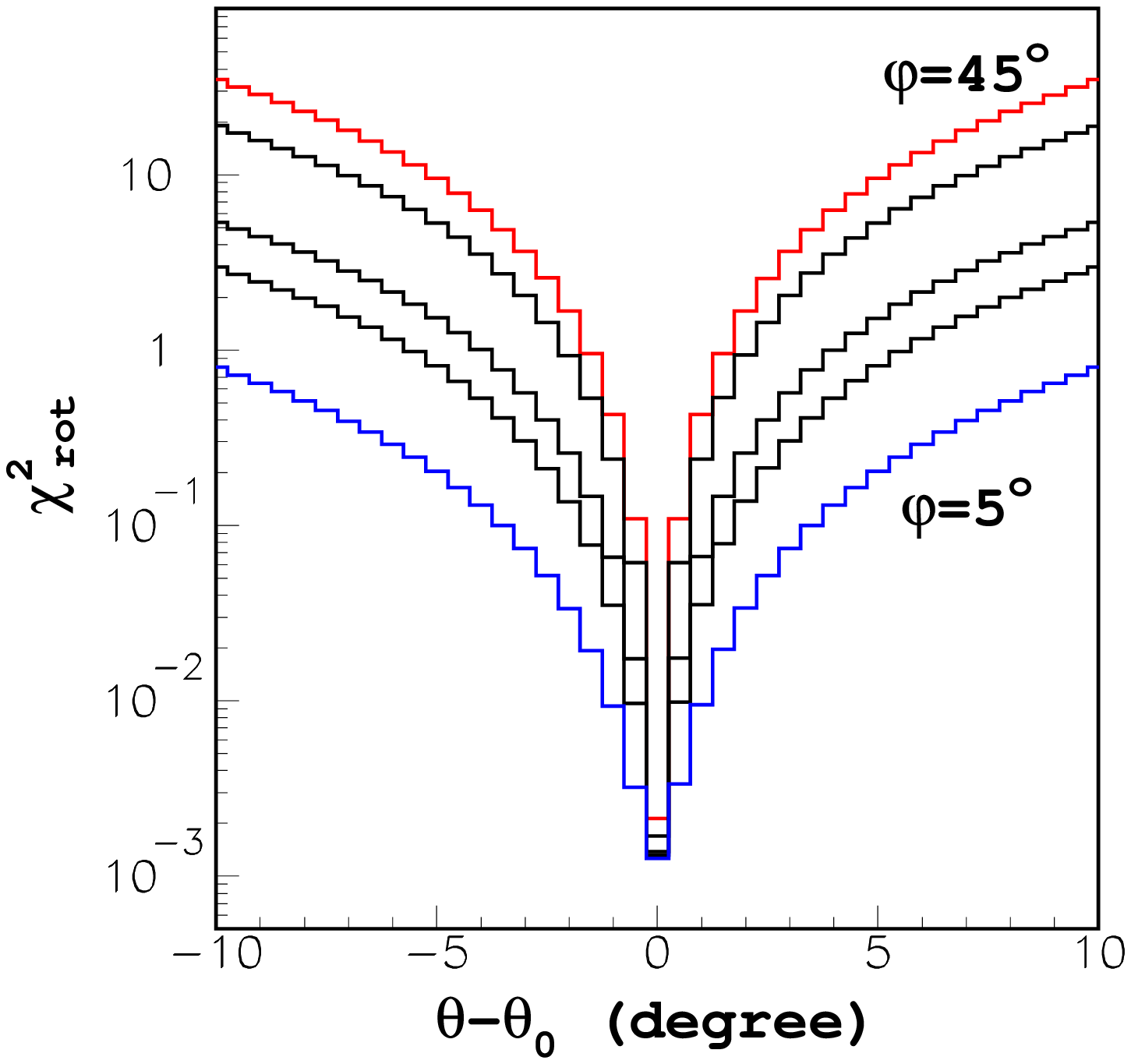}
\includegraphics[width=0.4\textwidth,trim=0.cm 0cm 0.cm 0.cm,clip]{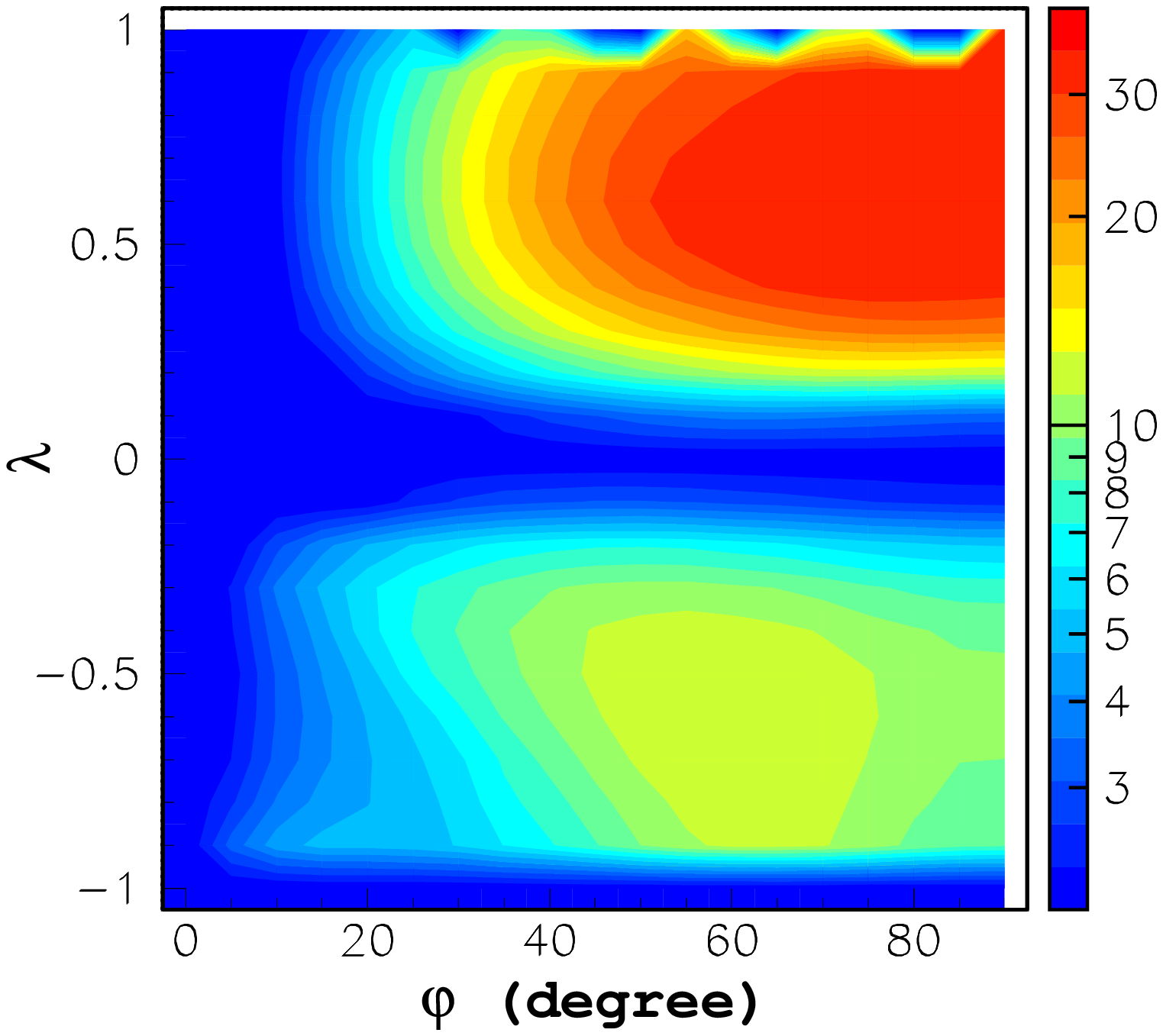}
\caption{Left: dependence of $\chi^2_{rot}$ (ordinate) on $\theta - \theta_0$ (degrees, abscissa) for $\lambda$=0.5 and $\varphi$=5$^\circ$, 10$^\circ$, 15$^\circ$, 30$^\circ$, 45$^\circ$ (from down up). Right: dependence of $S_{rot}$ on $\lambda$ (ordinate) and $\varphi$ (abscissa).}
\label{fig5}
\end{figure*}

In order to obtain some insight into this question, we consider a sample of uniformly distributed ($\lambda,\varphi,\psi$) triplets. Figure~\ref{fig3} (left) displays the values of $\sin\zeta$ at which an extremum of $V_z$ vs $z$ is found; they concentrate near $\sin\zeta=\pm$1. To see it more directly, we display in Figure \ref{fig4} a set of representative functions $\sin\zeta$ vs $V_z$. For $\sin\zeta=\pm$1,  $V_z=\pm V_0(1+\lambda \cos2\varphi)$ and when $\sin\zeta$ departs from $\pm$1, $V_z$ increases or decreases depending on the sign of $\lambda \sin2\varphi \cos\psi$. Globally, $V_z$ must increase with $z$; if it starts in the wrong direction, it needs to turn back and one obtains an extremum. When such extrema are in the vicinity of $\sin\zeta=\pm$1, they are not too harmful for de-projection: they simply mean that each end region of the Doppler velocity spectrum has to be assigned a broad range of $\sin\zeta$ values in the vicinity of the line of sight. However, the farther away they are from $\sin\zeta=\pm 1$, the larger the fraction of the Doppler velocity spectrum that becomes unsuitable for de-projection. In a majority of cases (68\%) there is no ``harmful'' extremum, defined as having $|\sin\zeta|>0.9$.\\

The right panel of Figure \ref{fig3} displays the region of the ($\lambda,\varphi$) plane
where no ``harmful'' extremum occurs, namely where $dV_z/dz$ does not cancel in the
$|\sin\zeta|<0.9$ region. In this region, as clearly illustrated in Figure \ref{fig4},
large values of $|V_z|$ are associated with large values of $|\sin\zeta|$ and
de-projection can be unambiguously performed as long as $V_0$ is large enough.
In the complementary ambiguous region, on the contrary, the larger values of $|V_z|$
are associated with intermediate values of $|\sin\zeta|$, the larger values of
$|\sin\zeta|$ being now associated with lower values of $|V_z|$: the Doppler
velocity spectrum is folded on itself, the larger values of $|V_z|$ being
associated with two values of $|\sin\zeta|$, making de-projection ambiguous
and unreliable. The strong qualitative difference between the two regions will
be seen to play a major role in the arguments developed in the present study.

\section{POSITION ANGLE OF THE PROJECTION OF THE STAR AXIS ON THE SKY PLANE}\label{sec.4}
The invariance of Relations \ref{eq.5} when $\psi$ changes sign implies an exact symmetry of the data-cube with respect to the plane of position angle $\theta$, perpendicular to the sky plane and containing the star axis and the line of sight. It is indeed a consequence of axi-symmetry and is expected to apply for any axi-symmetric model, not just the simple model used here for illustration. In principle, the symmetry plane, and therefore the value of $\theta$, can be simply found by minimizing the quantity
\begin{equation}\label{eq.8}
	\chi^2_{rot}=\sum [(f(x,y,V_z) - f(x^*,y^*,V_z)]^2/(\Delta f)^2 
\end{equation} 
where $(x^*,y^*)$ is the symmetric of $(x,y)$ with respect to direction $\theta$:	
\begin{equation}\label{eq.9}
	x^*=-(x \cos2\theta + y \sin2\theta ) \qquad		y^*=-(x \sin2\theta - y \cos2\theta)      
\end{equation}
and where $\Delta f$ is the uncertainty on the $f$ measurement. In the present section we use a simulation having a wind of the form given in Relation \ref{eq.3} and we simply take as $\Delta f$ the quadratic sum of the rms deviations of $f$ from its mean in the vicinity of each of ($x,y,V_z$) and ($x^*,y^*,V_z$).\\

As an illustration of the procedure, Figure \ref{fig5} (left) shows the dependence of $\chi ^2_{rot}$ on $\theta-\theta_0$ for $\lambda=0.5$ and various values of $\varphi$, with $\theta_0$ being the value of $\theta$ used to produce the simulated effective emissivity. The angle $\theta$ is undefined in two obvious cases: for an isotropic velocity distribution, $\lambda=0$, and for a star axis parallel to the line of sight, $\varphi=0$; the latter case is a trivial effect of geometry, the elementary solid angle being $d\Omega=\sin\varphi d\theta d\varphi$ rather than simply $d\theta d\varphi$. Indeed, the dependence of $\chi^2_{rot}$ on $\theta - \theta_0$ is observed to display a steep minimum at 0 as long as $\lambda$ and $\varphi$ are not too close from zero.\\

\begin{figure*}
\centering
\includegraphics[width=0.8\textwidth,trim=0.cm 0cm 0.cm 0.cm,clip]{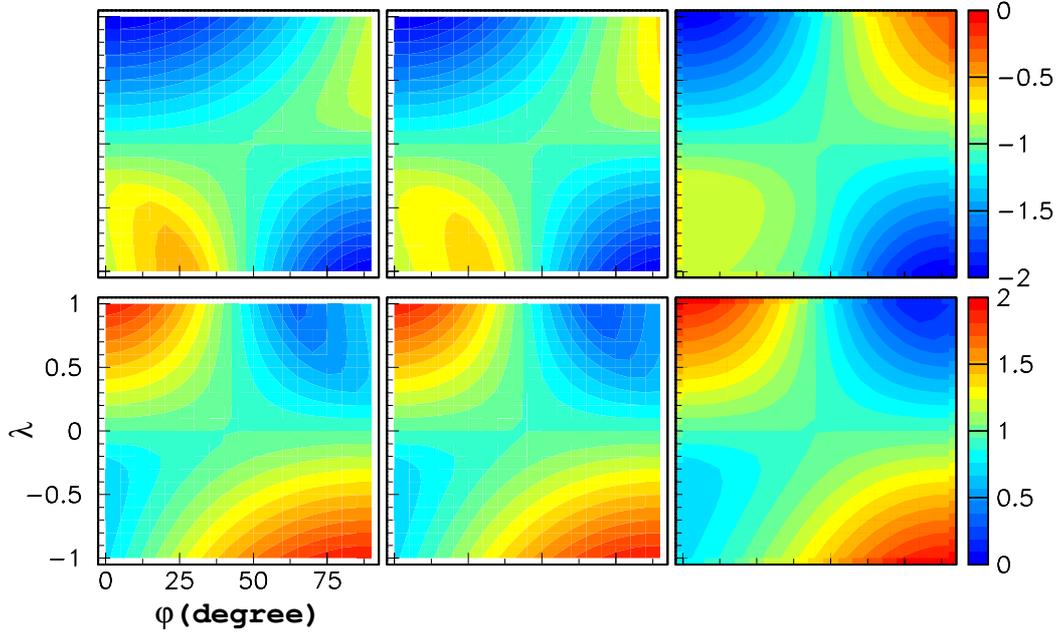}
\caption{Maps of $V_{zmin}/V_0$ (upper panels) and $V_{zmax}/V_0$ (lower panels) in the $\lambda$ (ordinate) vs $\varphi$ (abscissa) plane for $0<\psi<30^\circ$ (left), 30$^\circ<\psi<60^\circ$ (middle) and 60$^\circ<\psi<90^\circ$ (right).}
\label{fig6}
\end{figure*} 

In order to assess the accuracy of the $\theta$ measurement, we calculate the relative increase
$S_{rot}$ of $\chi^2_{rot}$ associated with a shift of $\pm 1^\circ$ from $\theta=\theta_0$:
$S_{rot}=1/2[\chi^2_{rot}(\theta=\theta_0 - 1^\circ)+\chi^2_{rot}(\theta=\theta_0 + 1^\circ)]/\chi^2_{rot}(\theta=\theta_0)$.
The larger $S_{rot}$ and the farther away from unity, the better defined is the value of
$\theta$. Figure \ref{fig5} (right) displays the dependence of $S_{rot}$ on $\lambda$
and $\varphi$. Its actual value when applied to real observations depends on the relevant
uncertainties, usually caused by the lumpiness of the effective emissivity rather than by
noise; but its behaviour in the ($\lambda,\varphi$) plane remains essentially the same as
found here. However, a systematic rather than random deviation from axi-symmetry may affect
the measurement of $\theta$, as was already discussed in the case of rotation by \citet{Diep2016}.
However, a systematic violation of axi-symmetry by the effective emissivity or by the wind
radial expansion velocity of the form $1 + \epsilon\cos(\omega-\omega_0)$ does not shift the
value of $\theta$ but simply broadens the minimum of $\chi^2_{rot}$, because, contrary to rotation,
it distorts in a same way the red-shifted and blue-shifted hemispheres.

\section{MAGNITUDE OF THE WIND VELOCITIES: WIDTH AND OFFSET OF THE DOPPLER VELOCITY SPECTRUM}\label{sec.5}
In general, the radial wind velocity $V$ is confined between two finite values; in the case
of the simple model, these are $V_{eq}$ and $V_{pole}$, respectively $V_0$($1-\lambda)$ and
$V_0$($1+\lambda$). However, at a given point ($x,y$) in the sky plane, the measured Doppler
velocity varies between two values $V_{zmin}$ and $V_{zmax}$ that are not simply related to the
above. Yet, when choosing a velocity distribution with which to de-project the effective
emissivity, it is essential to have some idea of its scale, meaning the value of $V_0$ in
the case of the simple model. In principle, having chosen a pair ($\lambda,\varphi$) for
de-projection, the values of $V_{zmin}/V_0$ and $V_{zmax}/V_0$ that they generate in each
pixel are known. One should then choose for $V_0$ the ratio between the measured values
of $V_{zmin}$ and $V_{zmax}$ and the model values of $V_{zmin}/V_0$ and $V_{zmax}/V_0$. The
values of $V_{zmin}/V_0$ and $V_{zmax}/V_0$ depend on the pixel.
In general, different pixels produce different values of $V_0$ and their mean, or better their maximum, should be
retained for de-projection. Note that changing $z$ in $-z$ and $\cos\psi$ in $-\cos\psi$
changes $V_z$ in $-V_z$ but leaves $dV_z/dz$ invariant; therefore it changes $V_{zmin}$
in $-V_{zmax}$ and $V_{zmax}$ in $-V_{zmin}$: it is sufficient to confine $\psi$ to the
[0$^\circ$,  90$^\circ$] interval. Figure \ref{fig6} displays the maps of $V_{zmin}/V_0$ and $V_{zmax}/V_0$
in the ($\lambda,\varphi$) plane for three intervals of $\psi$, each 30$^\circ$ wide,
covering between 0$^\circ$ and 90$^\circ$. Qualitatively, the dependence on $\psi$ is weak.
The main features reflect the effect of the width of the Doppler velocity spectrum,
which is large on the descending and small on the ascending diagonal. On the former,
from  ($\lambda,\varphi$)= ($+1,0^\circ$) to ($-1,90^\circ$), namely from a bipolar outflow
parallel to the line of sight to an equatorial outflow having its axis in the sky plane,
$V_{zmin}/V_0$ varies between $-1$ and $-2$ and $V_{zmax}/V_0$ between 1 and 2. On the latter,
from ($\lambda,\varphi$)=($-1,0^\circ$) to ($+1,90^\circ$), namely from a bipolar to an equatorial
outflow, both in the sky plane, $V_{zmin}/V_0$ varies between $-1$ and 0 and $V_{zmax}/V_0$ between 0 and 1.\\

\begin{figure*}
\centering
\includegraphics[width=0.8\textwidth,trim=0.cm 0.cm 0.cm 0.cm,clip]{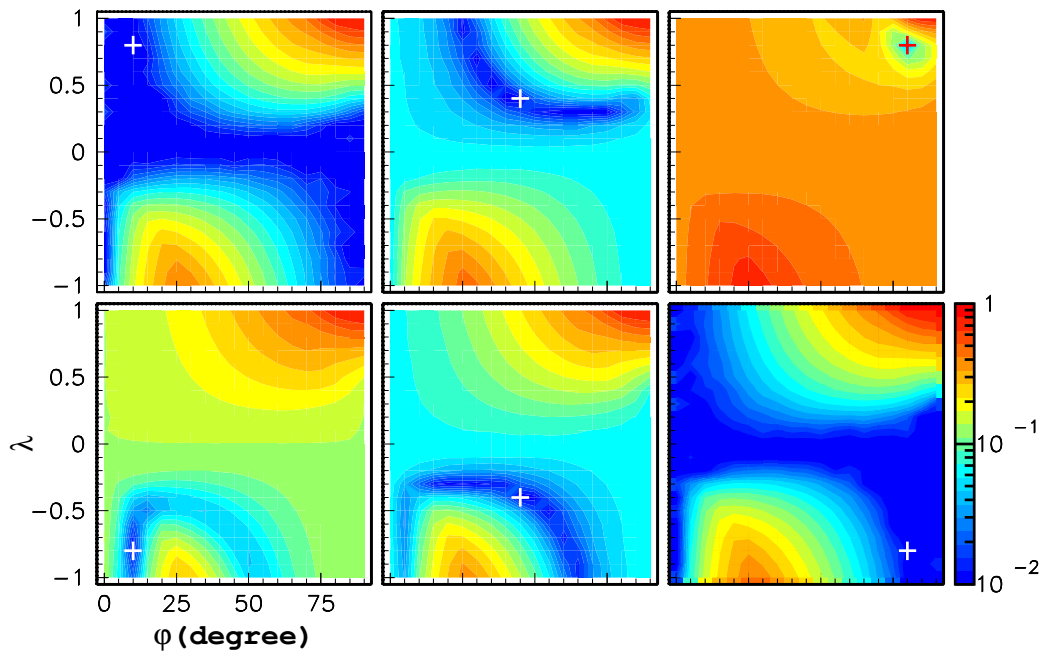}
\caption{Dependence of $Q$ on the values of $\lambda$ (ordinate) and $\varphi$ (abscissa) used in de-projection for a few simulated wind configurations of the form $V$(\kms)$=5(1-\lambda\cos2\alpha$) indicated as a cross. From left to right, the values of ($\lambda,\varphi$) used in the simulation are  ($\pm$0.8, 10$^\circ$), ($\pm$0.4, 45$^\circ$) and ($\pm$0.8, 80$^\circ$) respectively.}
\label{fig7}
\end{figure*}

When de-projecting the effective emissivity using a ($\lambda,\varphi$) pair of parameters,
agreement between the values obtained in each pixel for $V_0$ as a function of $\psi$ and $R$ is a
useful indicator of the suitability of the particular ($\lambda,\varphi$) pair to describe
the observations. The ratio $Q=V_{rms}/V_{mean}$ between their rms value $V_{rms}$ and their mean
value $V_{mean}$, calculated over the whole image, when too large, can be used to reject
unsuitable wind configurations: the study of the
dependence on position angle of the width and offset of the Doppler velocity spectrum is
not only a tool to obtain an evaluation of the scale of the space velocity (here $V_0$)
but also, in principle, to reject unsuitable values of the ($\lambda,\varphi$) pair. This
is illustrated in Figure \ref{fig7}, which displays the dependence of $Q$ on ($\lambda,\varphi$) for
a few typical simulated wind configurations. In all cases a steep minimum of $Q$ is obtained
in a narrow region of the ($\lambda,\varphi$) plane containing the values used in the simulation,
suggesting that an important fraction of the ($\lambda,\varphi$) plane could be eliminated
by simply requiring $Q$ not to exceed some threshold. In practice, however, as will be seen
in Section \ref{sec.7}, the minimum of $Q$ is much less steep for real than for simulated data.

We remark that if $V_0$ is slightly overestimated, its largest values will de-project in
a region of the observed Doppler velocity spectrum where there are no data and will accordingly
set the de-projected effective emissivity to zero. On the contrary, if $V_0$ is slightly
underestimated, the larger values of the observed Doppler velocity spectrum will be ignored.
It is therefore important, in practice, to make sure that the obtained value of $V_0$ is optimal
and, if necessary, to fine-tune it.

\section{MEASURING THE PROLATENESS PARAMETER $\lambda$ AND THE INCLINATION $\varphi$ OF THE STAR AXIS WITH RESPECT TO THE LINE OF SIGHT}\label{sec.6}
The results obtained in the preceding two sections are largely independent from the particular
form of the dependence of the effective emissivity on stellar latitude. They provide reliable
estimates of the scale $V_0$ of the space velocity distribution and of the position angle $\theta$
of the projection of the star axis on the sky plane. Moreover, they eliminate regions of the
($\lambda,\varphi$) plane that are unsuitable for de-projection. We are then left with two parameters,
$\lambda$ and $\varphi$, to be measured in the case of the simple model. In the general case,
several parameters will be necessary to describe the wind velocity in place of the pair ($V_0,\lambda$)
and $\lambda$ must be seen as a measure of the effective prolateness of the wind velocity distribution.\\

In the present section, we exploit the constraint resulting from the requirement of axi-symmetry
of the effective emissivity to help with the measurement of the ($\lambda,\varphi$) pair. This
constraint is less useful in some cases than in others. For example, in the case of a wind velocity
having its axis parallel to the line of sight any pair ($\lambda,\varphi$=0) used in de-projection
will produce an axi-symmetric effective emissivity, independently from the value assumed for $\lambda$.\\
 \begin{figure*}
   \centering
   \includegraphics[width=0.8\textwidth,trim=0.cm 0.cm 0.cm 0.cm,clip]{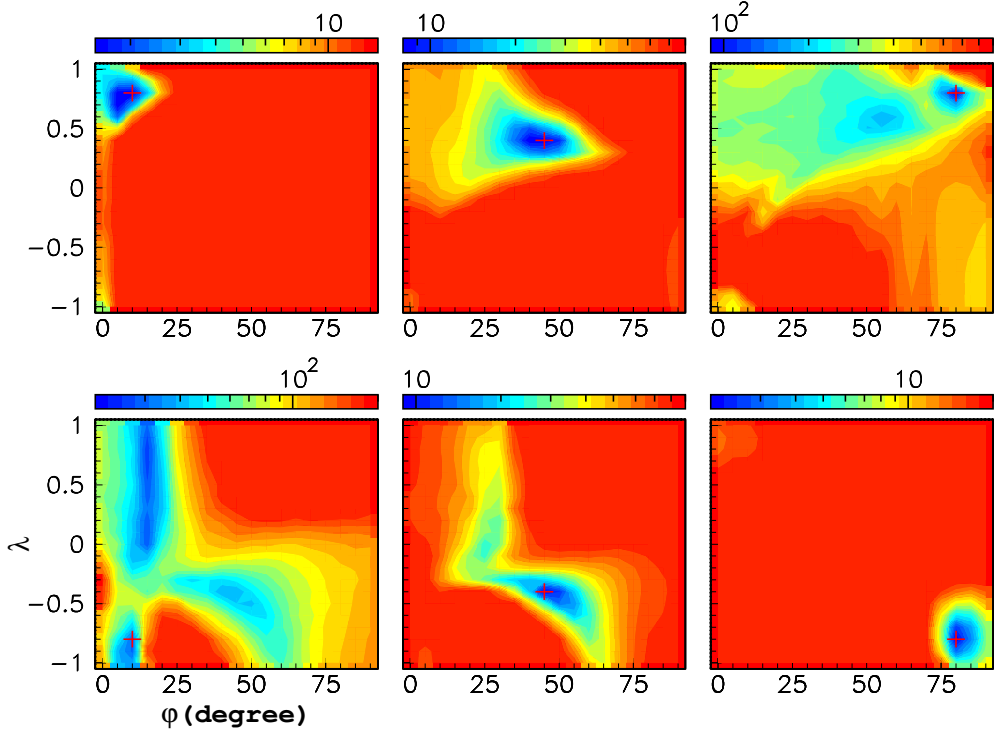}
   \caption{Dependence of $\chi^2_{axi}$ on the values of $\lambda$ (ordinate) and $\varphi$(abscissa) used in de-projection for the same simulated wind configurations indicated as a cross as in Figure \ref{fig7}. In each panel, the colour scale extends from minimum to ten times minimum.}
   \label{fig8}
   \end{figure*}

 In order to understand under which conditions the constraint of axi-symmetry is strong, we use a simple measure of the amount of axi-symmetry of the de-projected effective emissivity about axis $(\theta,\varphi)$: 
 \begin{equation}\label{eq.10}
	\chi^2_{axi}=\sum[\rho(r,\alpha,\omega)  - <\rho(r,\alpha)>]^2/(\Delta\rho)^2 
\end{equation}	
 where $\omega$ is the stellar longitude $(x'  = r\cos\alpha \cos\omega$, $y' = r\cos\alpha \sin\omega$
 and $\tan\omega = y'/x')$; $\rho (r,\alpha,\omega)$ is the effective emissivity de-projected using
 a wind configuration of axis $(\theta,\varphi)$, effective prolateness $\lambda$ and velocity scale
 $V_0$ estimated from the procedure described in the preceding section; \mbox{$<\rho(r,\alpha)>$} is its
 mean value at ($r,\alpha$), averaged over longitude $\omega$; $\Delta \rho$ is the uncertainty
 attached to the evaluation of $\rho$. The sum extends over the whole space over which measurements
 are available. The evaluation of $\Delta\rho=\Delta f |dV_z/dz|$, where $\Delta f$ is the uncertainty on the de-projected
 data-cube element, uses here the same estimate of  $\Delta f$ as used for $\chi^2_{rot}$ in Section \ref{sec.4}, namely the
 rms deviation of $f$ from its mean in the vicinity of the de-projected space point. In real cases,
 however, the definition of $\Delta f$ is delicate and needs to account for both experimental uncertainties
 and uncertainties attached to the de-projection; we discuss this point in more detail in Section \ref{sec.7}.
 In practice the calculation of $\chi^2_{axi}$ proceeds as follows: one chooses a circle having the
 star axis as axis and defined by its position ($R',z'$) in the star frame and scans over it by
 varying the stellar longitude $\omega$, each time calculating the space position ($x,y,z$) and the associated
 value of $V_z$ for the values of ($V_0,\lambda,\varphi$) of the axi-symmetric wind used in de-projection.
 
 The main weakness of this method is its mishandling of cases where two different values of $z/r$
 are associated with a same bin of Doppler velocity. Such bins contain contributions from each of
 the two regions but the de-projection algorithm wrongly assigns their total content to each of
 the two regions, generating double-counting. As was already remarked in Section \ref{sec.3}, one does not
 know how to share it between the two regions and such bins are unsuitable for de-projection.
 We have seen in Section \ref{sec.3} that for $z/r=\pm$1, $V_z=\pm V_0(1 + \lambda \cos2\varphi)$ and that
 when $z/r$ departs from $\pm$1, $V_z $ increases or decreases depending on the sign of
 $\lambda \sin2\varphi \cos\psi$. Rather than simply ignoring the ambiguous intervals of Doppler
 velocity between $V_{zmin}$ and $-V_0(1+\lambda\cos2\varphi)>V_{zmin}$ and between
 $V_0(1+\lambda\cos2\varphi)<V_{zmax}$ and $V_{zmax}$, we simply assign to each of the two associated
 $z$ values half the de-projected emissivity, thereby avoiding double-counting.

 Figure \ref{fig8} illustrates the dependence of $\chi^2_{axi} $ on parameters $\lambda$ and $\varphi$
 for the same simulated wind configurations as displayed in Figure \ref{fig7}. The effective emissivity
 used to produce the simulated data-cube is of the form 1/$r^2$ for $0.5<r<5$ arcsec with no dependence on
 stellar latitude in order not to bias the evaluation of $\chi^2_{axi}$ or, more precisely, not to complicate
 its interpretation. For $r>5$ arcsec, the effective emissivity is taken to cancel and for $r<0.5$ arcsec  it is
 taken to be constant.
 
 In all cases $\chi^2_{axi}$ is minimal at the simulated value of ($\lambda,\varphi$) but in some cases
 a broader region of the ($\lambda,\varphi$) plane is observed to be equally acceptable. The minimum
 is better behaved along the descending diagonal (wind velocities out of the sky plane) than along
 the ascending diagonal (wind velocities near the sky plane): in the former case $\chi^2_{axi}$ is
 more efficient than $Q$ to constrain the ($\lambda,\varphi$) pair while in the latter case $Q$ is more
 efficient than $\chi^2_{axi}$.\\
 \begin{figure*}
 \centering
 \includegraphics[width=0.8\textwidth,trim=0.cm 0.cm 0.cm 0.cm,clip]{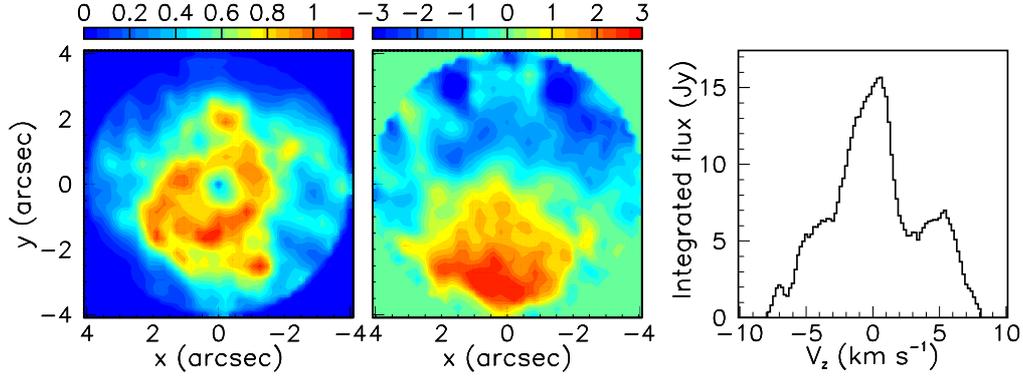}
 \caption{NOEMA observations of the CO(2-1) emission of RS Cnc in the region $R<4$ arcsec. Left: sky map of the intensity (Jy~beam~$^{-1})$ multiplied by $R$ (arcsec). Middle: sky map of $<V_z>$ (km s$^{-1}$). Right: Doppler velocity $V_z$ spectrum (km s$^{-1})$ integrated over the region $R<4$ arcsec. In the first two panels north is up and east is left.}
 \label{fig9}
 \end{figure*}

 Using $\varphi=0$ for de-projection deserves a special comment; in this case, a given circle ($R',z'$)
 is made of points having all the same value of $z=z'$ and the same value of $R=R'$, therefore a same
 value of $V_z$ and a same value of $dV_z/dz$: $\chi^2_{axi}$ is simply a measure of the axi-symmetry
 of the data-cube about the $z$ axis for Doppler velocities $|V_z|<(1+\lambda)V_0$. In particular,
 when the real wind and the wind used for de-projection are both axi-symmetric with respect to the
 $z$ axis, with respective prolateness $\lambda_{true}$ and $\lambda_{deproj}$, $\chi^2_{axi}=0$ is of
 no help to constrain the value of $\lambda_{true}$. The same is true of the study of $V_{zmin}$,
 $V_{zmax}$ and $Q$, in the approximation where the wind velocity and mass loss rate do not depend
 on $r$: whatever the value of $\lambda$ the extrema of the velocity spectrum are reached in all
 pixels at a same value of $V_{zmax}$ and $V_{zmin}=-V_{zmax}$. The result of the de-projection depends
 on the location of $\lambda_{true}$ and $\lambda_{deproj}$ with respect to the value $-0.2$ that
 separates ambiguous from unambiguous cases. When both $\lambda_{true}$ and $\lambda_{deproj}$ are
 larger than $-0.2$, the maximum of $|V_z|$ is reached at the poles and the value of $V_0$ used
 for de-projection is the true value multiplied by $(1+\lambda_{true})/(1+\lambda_{deproj})$.  However,
 when one of the $\lambda$ parameters takes values that vary between $-0.2$ and $-1$, the maximum
 of $|V_z|$ is reached at values of $|\sin\zeta|$ that vary between 1 and $1/\sqrt{3}$, the
 corresponding values of $|V_z|$ varying between $0.8V_0$ and $4V_0/[3\sqrt{3}]\sim 0.77V_0$
 with a minimum at $\lambda=-1/2$ where $|\sin\zeta|=1/\sqrt{2}$ and $|V_z|=1/\sqrt{2}V_0\sim 0.71V_0$.
 This near independence on $\lambda$ ($\sim \pm 3V_0/4$) of $V_{zmin}$ and $V_{zmax}$ when $\varphi=0$
 and $\lambda<-0.2$ could already be seen in Figure \ref{fig6}.
 In general, when using $\varphi=0$ and $\lambda<-0.2$ for de-projection of a ($\lambda_0,\varphi_0$) data-cube
 located in the unambiguous region will require the use of a $V_0$ value significantly larger than the true value
 and will deproject the large values of $|V_z|$, associated with large values of $|\sin\zeta|$, to intermediate
 values of $|\sin\zeta|$ resulting in relatively low values of $\chi^2_{axi}$.
\section{CASE STUDIES: RS C\lowercase{NC} AND EP A\lowercase{QR}} \label{sec.7}
In the preceding sections, arguments were developed using simulated rather than real observations. In practice, the morphology of circumstellar envelopes of evolved stars are far from being as smooth and well-behaved as those simulated in the present study. The question of the practicability of using the arguments and the tools developed in the preceding sections remains therefore open at this stage. While each particular case must be considered separately, we find it useful to devote the present section to two case studies in order to get some idea of the nature and magnitude of the difficulties that one may have to face. \\
\begin{figure*}
\centering
\includegraphics[width=0.8\textwidth,trim=0.cm 0.cm 0.cm 0.cm,clip]{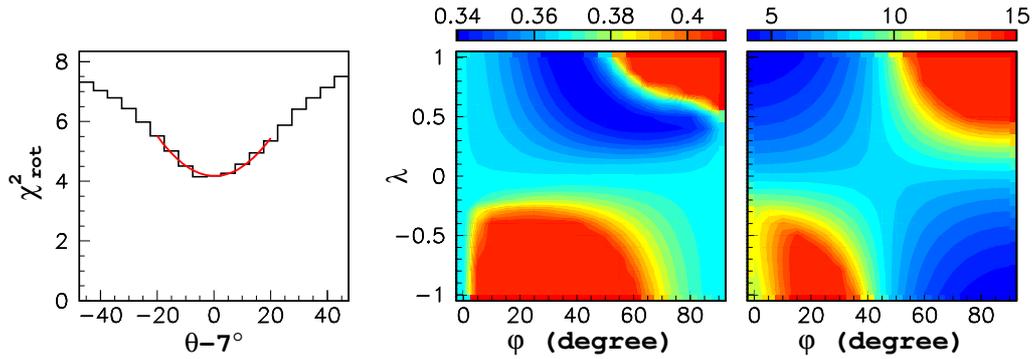}
\caption{RS Cnc CO(2-1) emission. Left: dependence of $\chi^2_{rot}$ on $\theta - 7^\circ$; the curve is a
  parabolic fit.  Middle: ($\lambda,\varphi$) map of $Q$ for the region $R < 4$ arcsec. The colour scale runs
  from minimum ($\sim$ 0.34) to 1.2 times minimum. Right: ($\lambda,\varphi$) map of the velocity scale $V_0$ (\kms).}
\label{fig10}
\end{figure*}
\subsection{RS Cnc}
The first star selected for this purpose is RS Cnc, an S-type AGB star (CSS 589 in Stephenson's
1984 catalogue) in the thermally-pulsing phase \citep[]{Lebzelter1999}, which can be considered as
representative of its family, the morphology of its envelope being quite clumpy but not excessively.
Analyses of Plateau de Bure observations of its CO(1-0) and CO(2-1) emissions have been published
earlier \citep[]{Libert2010, Hoai2014, Nhung2015a, Le Bertre2016}. Here, we use CO(2-1) observations
from the upgraded Plateau de Bure interferometer (NOEMA) having a beam size (FWHM) of $0.44\times0.28$
arcsec$^2$ and a spectral resolution of $\sim$ 0.2 km s$^{-1}$. The noise level per data cube element
is 1.6 mJy beam$^{-1}$ for pixels of $0.07\times 0.07$ arcsec$^2$. In order to improve the signal to noise
ratio, we limit the study to the region $R<4$ arcsec and we group pixels by $3\times3=9$, meaning
$0.21\times 0.21$ arcsec$^2$; we use as experimental uncertainty on the brightness the quadratic
sum of the noise and of 20\% of the measured value. Figure~\ref{fig9} displays sky maps of the measured
intensity multiplied by $R$ and of the mean Doppler velocity, together with the integrated Doppler velocity spectrum.\\

As already apparent from the map of the mean Doppler velocity, the projection of the star axis on
the sky plane is nearly north-south oriented. Indeed we find that $\chi^2_{rot}$ is minimal for
$\theta_0$ $\sim$ 7$^\circ$ and we display in Figure \ref{fig10} (left) the dependence on $\theta-\theta_0$
of its normalized value (divided by the number of degrees of freedom).  The value of $\chi^2_{rot}$
at minimum, $\sim$ 4, is the result of the relative lumpiness of the measured brightness. The uncertainty
on $\theta$ is accordingly poorly defined; as an indication of how well $\theta$ is measured we
quote as effective uncertainty the value associated with a 10\%  increase of $\chi^2_{rot}$ with
respect to minimum, namely $\theta$ =7$^\circ\pm$10$^\circ$. Indeed, masking measured brightness values
below noise level (one sigma) minimizes $\chi^2_{rot}$ at $\theta \sim$ 10$^\circ$ instead of $\sim$ 7$^\circ$.

Having obtained an evaluation of  $\theta$ we rotate the data-cube by $7^\circ$ about the line of sight in order
to have effectively $\theta=0$ and proceed with the evaluation of the scale $V_0$ of the wind velocity and the
exclusion from the ($\lambda,\varphi$) plane of regions unsuited for de-projection as was done in Section~\ref{sec.5}
using simulated data. The dependence of $Q$ on $\lambda$ and $\varphi$ is displayed in the central panel of
Figure \ref{fig10}. At first sight, the resemblance with the upper-middle panel of Figure \ref{fig7}
($\lambda=0.4$, $\varphi=45^\circ$) is striking; however, the minimum is now considerably less steep than
it was for simulated data and extends over a broad region that excludes wind velocities closely confined
to the sky plane, whether bipolar or equatorial outflows. The resulting $V_0$ values are displayed in the
right panel of Figure \ref{fig10} and range between $\sim$ 4 and $\sim$ 15 km~s$^{-1}$, the lower values being associated with
prolate bipolar outflows near the line of sight or oblate equatorial outflows having their axis near the
sky plane, namely wind configurations in the unambiguous de-projection region. The larger values are instead
confined to the ambiguous de-projection region of the ($\lambda,\varphi$) plane.

\begin{figure*}
\centering
\includegraphics[height=8cm,trim=0.cm 0cm 1.5cm 0.cm,clip]{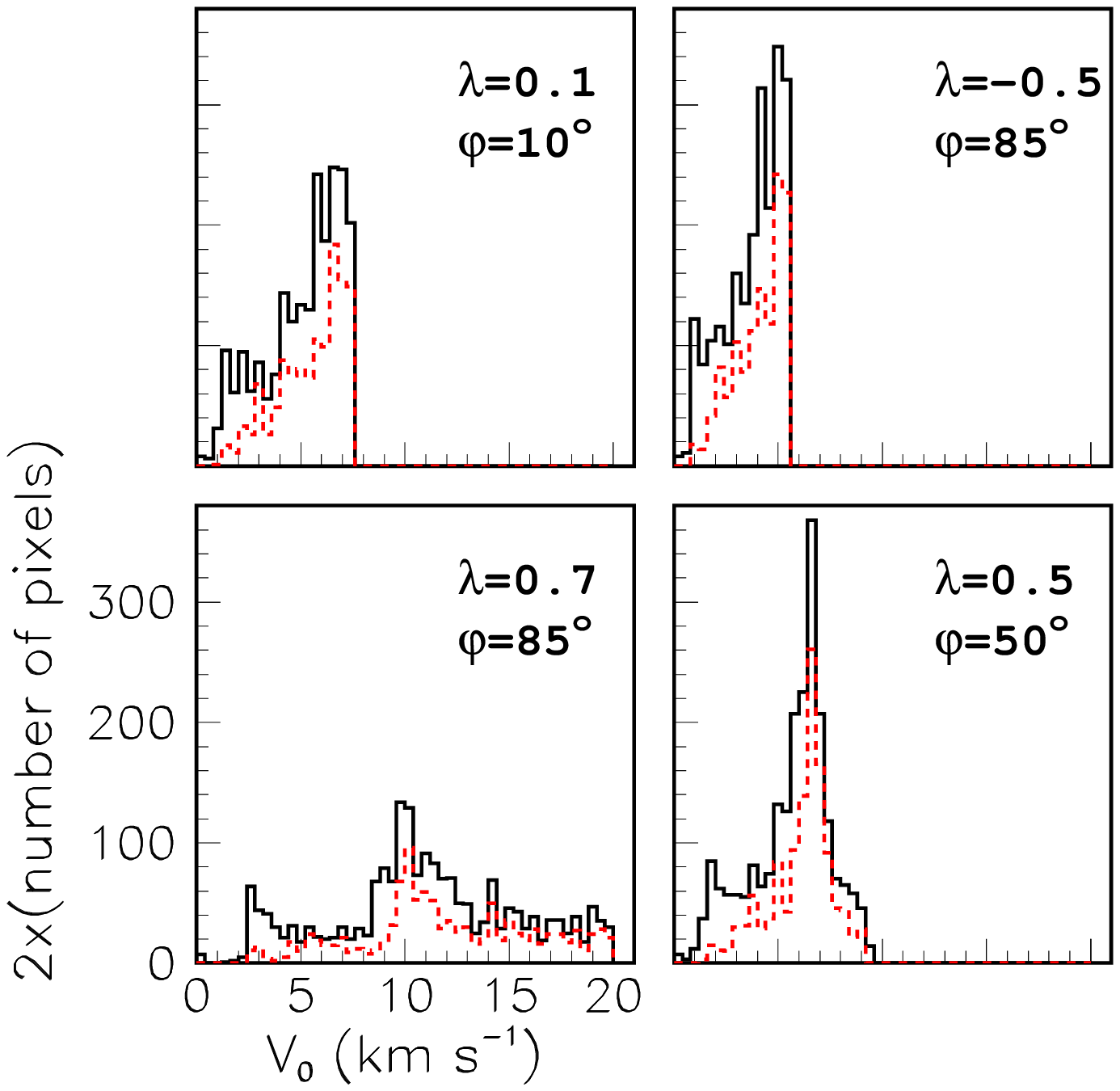}
\includegraphics[height=8cm,trim=.9cm 0cm 0.cm 0.cm,clip]{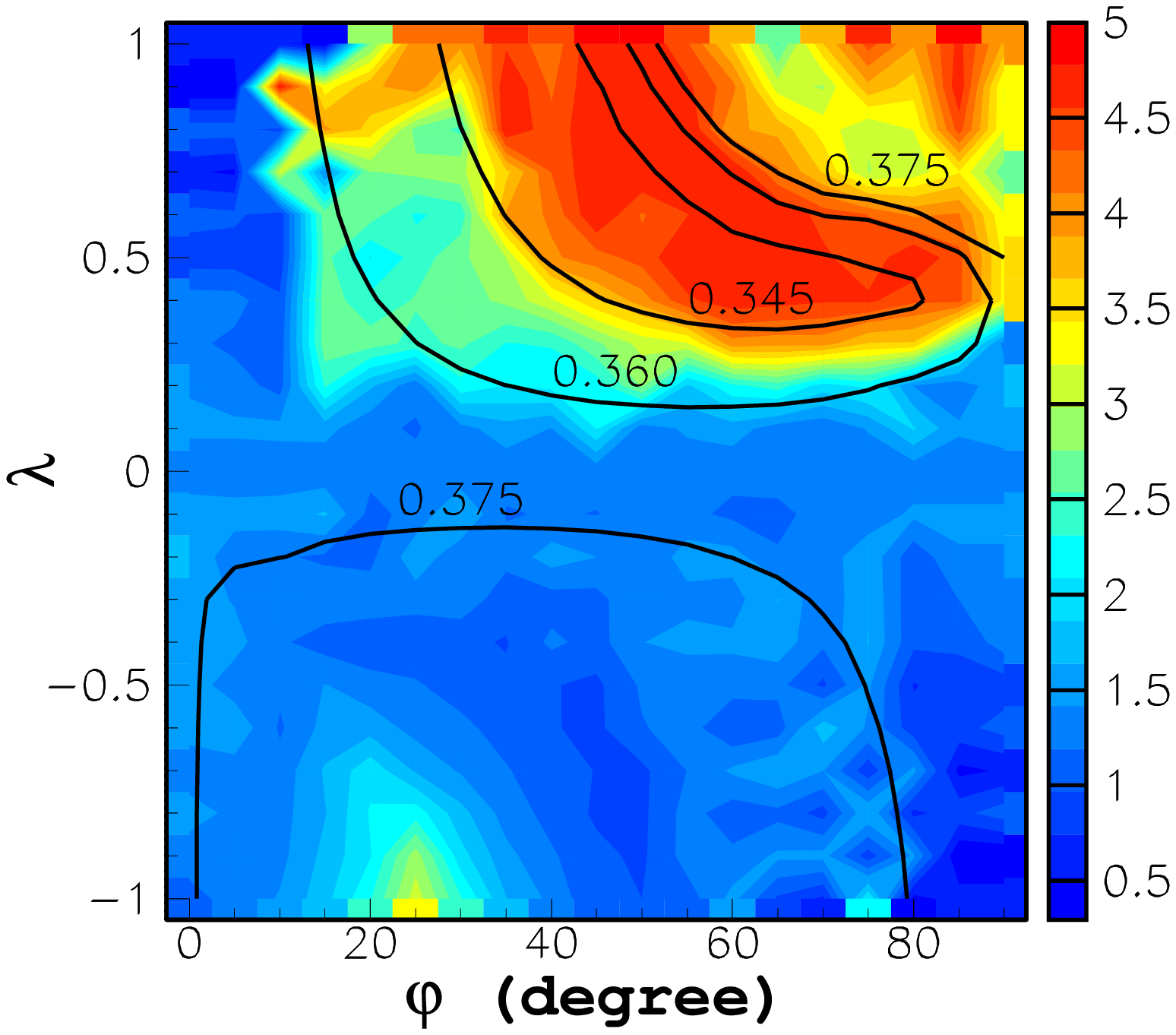}
\caption{RS Cnc. Left: $V_0$ distributions obtained for four different values of the ($\lambda,\varphi$) pair and for
  $R<4$ arcsec (solid) or $R<3$ arcsec (dashed). Right: distribution of $P/T$ in the ($\lambda,\varphi$) plane for
  $R<3$ arcsec. The contours correspond to $Q=0.345$, 0.360 and 0.375 respectively. }
\label{fig11}
\end{figure*}

A major cause of error in the evaluation of $V_0$ is the difficulty to measure accurately $V_{zmin}$ and/or $V_{zmax}$
in pixels where the brightness is close to noise level at the extremities of the Doppler velocity spectrum. In
calculating in each pixel the values of $V_{zmin}$ and $V_{zmax}$ we used brightness in excess of two noise $\sigma$'s
and checked that the result was essentially unaffected when using instead one or three noise $\sigma$'s. The left
panels of Figure \ref{fig11} display distributions of the values obtained for $V_0$ in each pixel (two values per pixel)
for some representative values of the ($\lambda,\varphi$) pair. Ideally, if the selected wind configuration
describes the data well, the $V_0$ distribution must have the shape of a narrow peak centered at the proper $V_0$
value. Such a peak is indeed visible for some values of the ($\lambda,\varphi$) pairs, but it is broad and
accompanied by a low $V_0$ tail: exploiting the information contained in the dependence over the sky plane
of $V_{zmin}$ and $V_{zmax}$ is clearly more difficult when dealing with real data than it is with simulated
data. In pixels where the brightness is close to noise level at the extremities of the Doppler velocity spectrum
$|V_{zmin}|$ and/or $|V_{zmax}|$ are under-evaluated and the same is therefore true for $V_0$. Pixels at larger
values of $R$, being associated with lower intensities, are more likely to be of that kind. Indeed, as can
be seen in Figure \ref{fig11}, the distribution of $V_0$ obtained inside the circle $R<3$ arcsec is narrower
and less contaminated by a low $V_0$ tail than inside the circle $R<4$ arcsec. The distributions displayed
in the left panels of Figure \ref{fig11} suggest that a figure of merit revealing the presence of a peak in
the $V_0$ distribution might be more efficient than $Q$ in exploiting the information carried by $V_{zmin}$ and $V_{zmax}$.
We use a simple algorithm to evaluate, for each value of the ($\lambda,\varphi$) pair the peak to tail ratio
of the $V_0$ distribution, $P/T$. The distribution of $P/T$ over the ($\lambda,\varphi$) plane is displayed
in Figure \ref{fig11} (right). It is consistent with the information carried by $Q$ but is much more selective:
a reasonable cut $Q>0.36$ is only a factor 1.06 above minimum while an equivalent cut $P/T<2$ is a factor $\sim$ 4
above minimum and a factor 2.5 below maximum, allowing for safely rejecting regions of the ($\lambda,\varphi$) plane having $P/T<2$.

\begin{figure*}
\centering
\includegraphics[height=5.6cm,trim=0.5cm 0.cm 0.5cm 0.cm,clip]{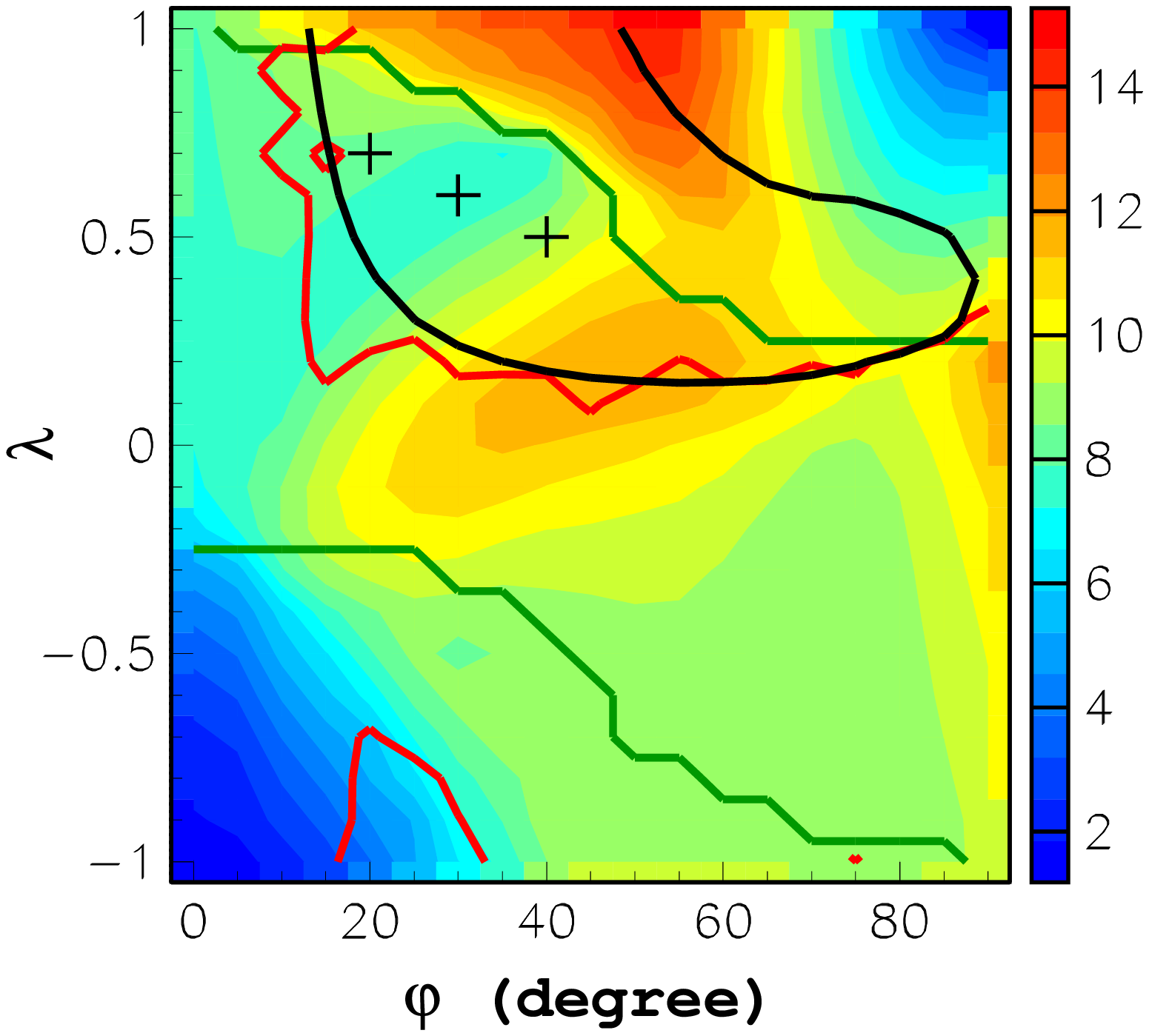}
\includegraphics[height=5.6cm,trim=0.cm 0.cm .9cm 0.cm,clip]{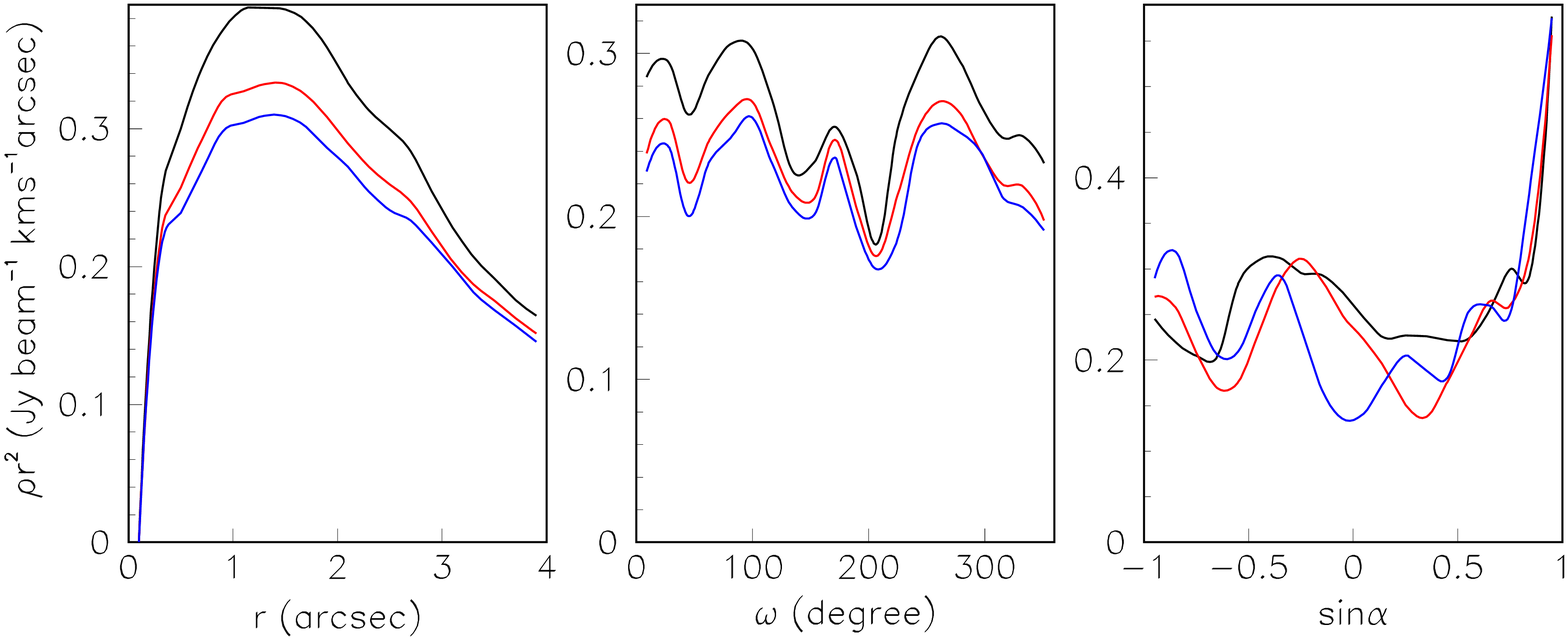}
\caption{RS Cnc. Left: distribution of $\chi^2_{axi}$ in the ($\lambda,\varphi$) plane. The contours are for ambiguous
  de-projection (green), $Q=0.36$ (black) and $P/T=2$ (red). Rightmost panels (from left to right): dependence
  of the de-projected emissivity multiplied by $r^2$ (Jy beam$^{-1}$\kms arcsec) on $r$ (arcsec), $\omega$ (degree)
  and $\sin\alpha$ respectively. The values of ($\lambda,\varphi$) are, from up down, (0.5, 40$^\circ$), (0.6, 30$^\circ$) and (0.7, 20$^\circ$), corresponding to the crosses in the left panel.}
\label{fig12}
\end{figure*}

In order to further restrict the acceptable region of the ($\lambda,\varphi$) plane, we still need to exploit
the constraint of axi-symmetry as was done in Section \ref{sec.6} using simulated observations. This, however, cannot
be done reliably in the regions of ambiguous de-projection (right panel of Figure \ref{fig3}). As these have a large
overlap with the regions disfavoured by the analysis of the ($\lambda,\varphi$)-dependence of $V_{zmin}$ and
$V_{zmax}$, we may exclude them from the analysis by rejecting regions containing a ``harmful'' extremum of
the $V_z$ vs $z$ relation as defined in the right panel of Figure \ref{fig3}, namely regions where $dV_z/d_z$
cancels in the $|\sin\zeta| < 0.9$ interval. When an extremum occurs at $|\sin \zeta| > 0.9$, we share
the de-projected emissivity equally between the two associated values of $\sin\zeta$. The distribution of
$\chi^2_{axi}$ in the ($\lambda,\varphi$) plane is displayed in Figure \ref{fig12} together with the boundaries
associated with regions containing a ``harmful'' extremum and with the contours associated with $Q=0.36$ and $P/T=2$.
The main contribution of $\chi^2_{axi}$ is to disfavour wind configurations close to spherical with an axis
at intermediate inclination with respect to the sky plane, leaving a relatively narrow region of acceptable
bipolar outflows in the upper-left quadrant of the ($\lambda,\varphi$) plane with $\lambda$ between $\sim$ 0.3
and $\sim$ 0.8 and $\varphi$ between $\sim 15^\circ$ and $\sim 45^\circ$.  

The joint analysis of CO(1-0) and CO(2-1) emissions presented in \cite{Nhung2015a} gave
$(V_0, \lambda, \theta, \varphi)=(5.0\pm0.2$ \kms, 0.50$\pm$0.02, 9$^\circ\pm 6^\circ$, 38$^\circ\pm 2^\circ$) ,
however allowing for some r-dependence of the expansion velocity. This corresponds indeed to values of
$Q<0.36$, $P/T>2$ and $\chi^2_{axi}<10$ favoured by the present study. We display in Figure \ref{fig12}
(right) the dependence of the de-projected emissivity, multiplied by $r^2$ on respectively $r$, $\omega$
and $z'/r=\sin\alpha$ for three different values of the ($\lambda,\varphi$) pair in the favoured region,
($\lambda,\varphi$)= (0.5, 40$^\circ$), (0.6, 30$^\circ$) and (0.7, 20$^\circ$) respectively. All three distributions
are averaged in the sphere $r<4$ arcsec. A remarkable result is the independence of the $r$-dependence of
the de-projected effective emissivity on the value of ($V_0, \lambda, \varphi$) used for de-projection,
a result that had been anticipated and explained in Section~\ref{sec.2} using Relations~\ref{eq.5}. However, the longitudinal
dependence differs significantly from uniform, with an excess in the $180^\circ<\omega<360^\circ$ hemisphere
compared with $0<\omega<180^\circ$. The latitudinal dependence displays a clear asymmetry with respect to
the star equator.  Table~\ref{tab1} summarizes the results.

Commenting further on these results goes beyond the scope of the present study. Physics arguments need now
to be used for interpreting the observed behaviour of the de-projected emissivity and its relation with the
probably much too simple form assumed for the wind configuration. Yet, the present results are of considerable
help in constructing a physics model of the morphology and kinematics of the circumstellar envelope and have
provided a deep insight in the constraints that such a model has to obey.

\begin{table*}
  \centering  
  \caption{ 
 De-projection of RS Cnc CO(2-1) observations: summary of results }    
  \label{tab1}        
  \renewcommand{\arraystretch}{1.1}
  \begin{tabular}{|c|c|c|c|c|c|c|c|}
    
    \hline Case              &$V_0$ (km /s)&$\lambda$&$\varphi(^\circ) $ & $\theta(^\circ) $ & $Q$  & $P/T$ &$\chi^2_{axi} $ \\ 
    
    \hline \cite{Nhung2015a} & 5.0$\pm$0.2&0.50$\pm$0.02     &	38$\pm$2&9$\pm$6	&$-$          &$-$        &$-$   \\
    \hline 1                &6.0         & 0.5              & 40       & 7             &0.344        &3.6        &9.5\\
    \hline 2                 &5.0         &0.6               &30        &7              &0.345        &2.0        &7.5\\
    \hline 3                 &4.5         &0.7               &20        &7              & 0.356       &2.2        &8.7\\
     
    \hline
  \end{tabular} 
\end{table*}

\begin{figure*}
  \centering
  \includegraphics[width=0.6\textwidth,trim=0.cm 2.2cm 1.cm 1.9cm,clip]{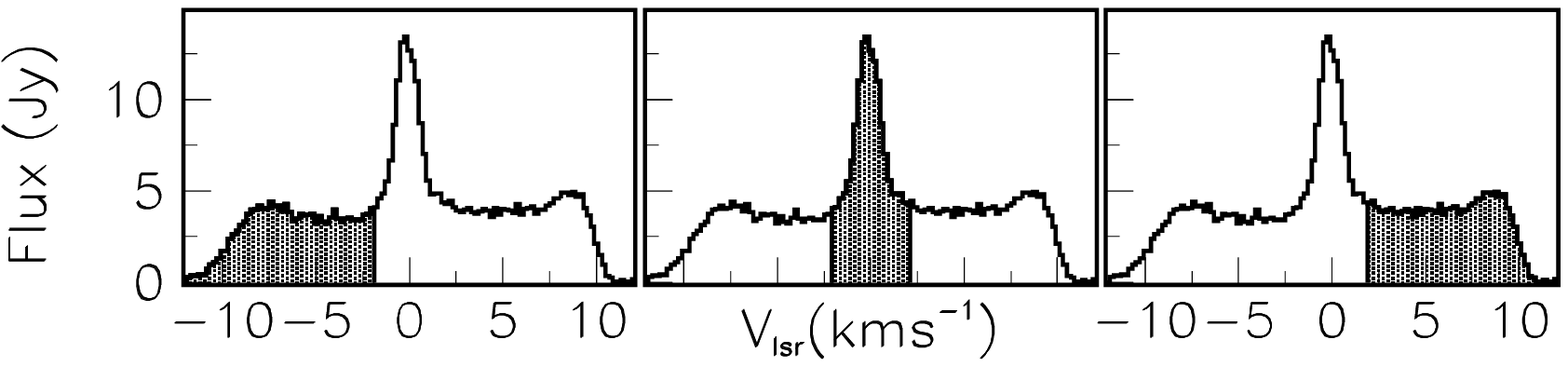}
  \includegraphics[width=0.6\textwidth,trim=0.cm 2.2cm 1.cm 1.7cm,clip]{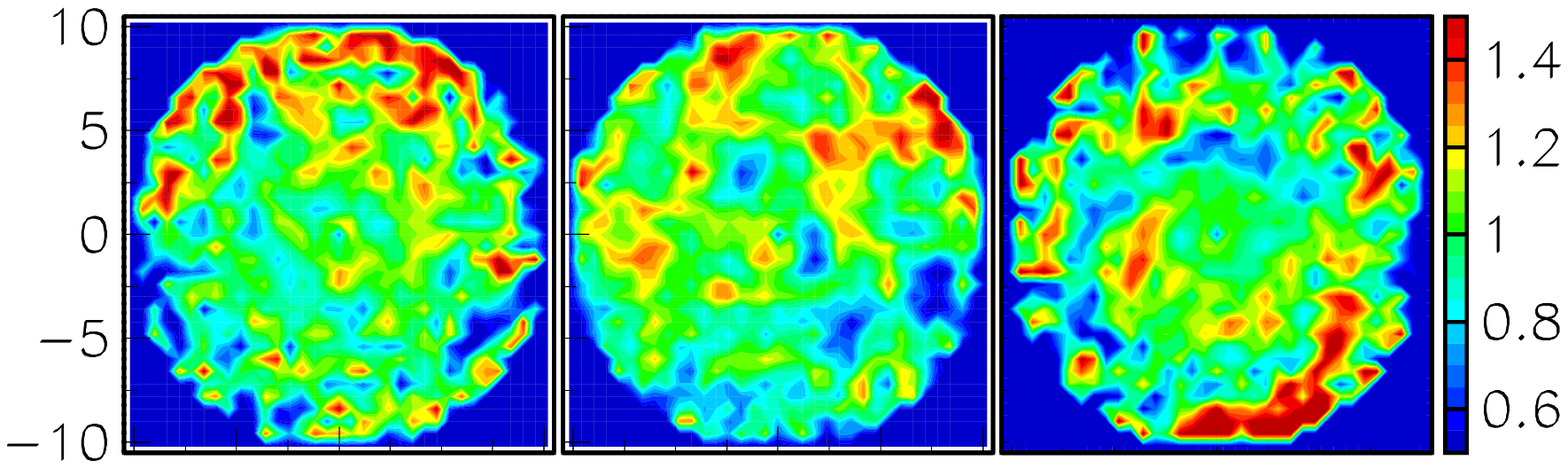}
  \includegraphics[width=0.6\textwidth,trim=0.cm 2.2cm 1.cm 1.7cm,clip]{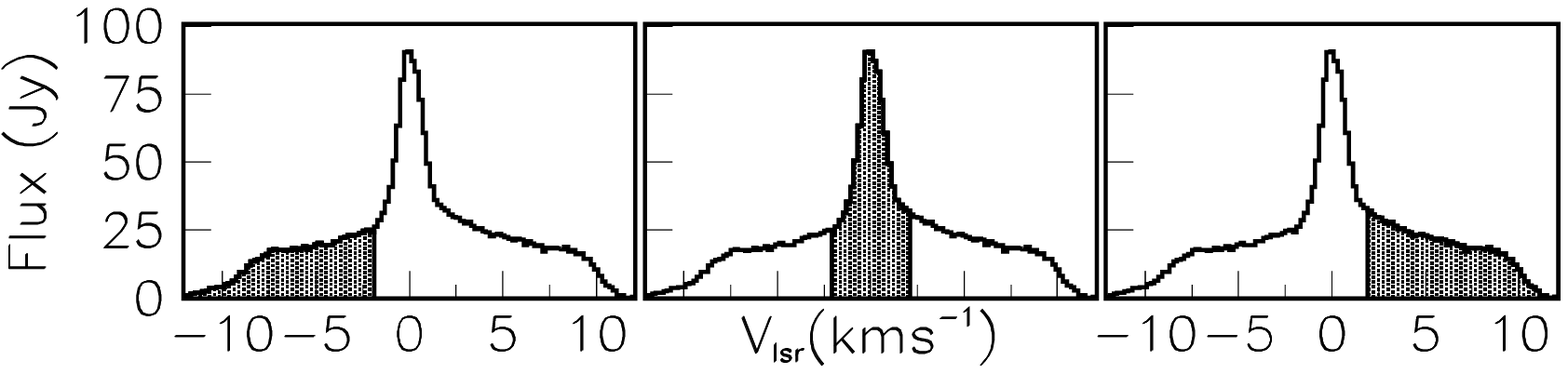}
  \includegraphics[width=0.6\textwidth,trim=0.cm 1.5cm 1.cm 1.7cm,clip]{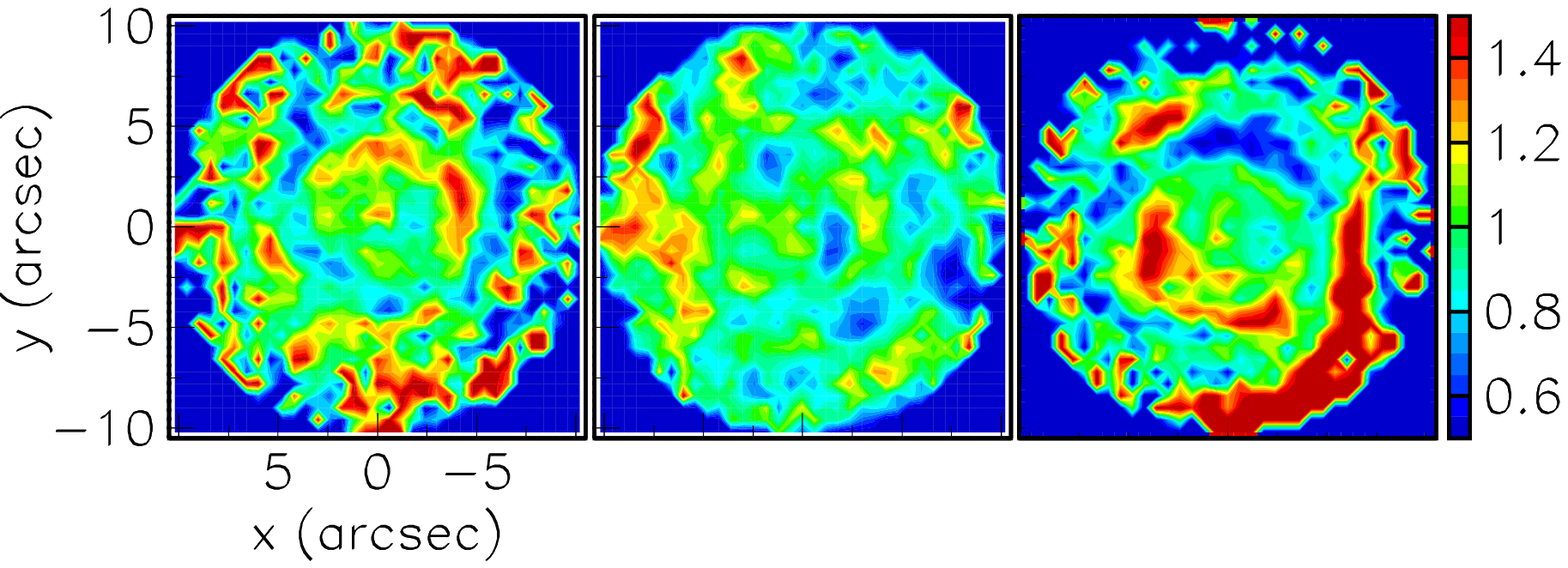}
  \caption{EP Aqr CO(1-0) (upper panels) and CO(2-1) (lower panels) emission. Sky maps of $\Delta(x,y)-1$ for $R<10$ arcsec. Left panels are for $V_z<-2$ \kms, central panels for $|V_z|<2$ \kms\ and right panels for $V_z>2$ \kms\ as indicated in the Doppler velocity spectra displayed above the sky maps. From \citet{Nhung2018}.}
\label{fig13}
\end{figure*}

\subsection{EP Aqr}
The second star used for illustration is EP Aqr, an oxygen-rich M type AGB star that is probably at the beginning of its evolution on the thermally pulsing phase \citep[]{Lebzelter1999, Cami2000} in spite of observations of trailing gas \citep[]{Cox2012, Le Bertre2004} suggesting a mass loss episode at the scale of $10^4$ to $10^5$ years. Recently, observations of $^{12}$CO(1-0) and $^{12}$CO(2-1) emissions using the IRAM 30-m telescope and the Plateau de Bure Interferometer have been reported \citep[]{Winters2003,Winters2007, Nhung2015b,Le Bertre2016}. Here, we use CO(1-0) and CO(2-1) observations made in Cycle 4 of ALMA operation \citep[2016.1.00026.S]{ Nhung2018}. The beam size (FWHM) is respectively $0.78\times0.70$   and $0.33\times0.30$ arcsec$^2$ and the noise respectively 8 and 7 mJy beam$^{-1} $ for pixels of respectively $0.2\times 0.2$ and $0.1\times0.1$ arcsec$^2$ and Doppler velocity bins of 0.2 \kms. Similar but different ALMA observations, including also SiO and SO$_2$ emission, have recently been presented by \cite{Homan2018}.

\begin{figure*}
\centering
\includegraphics[height=5cm,trim=0.5cm 0.7cm 0.9cm 1.2cm,clip]{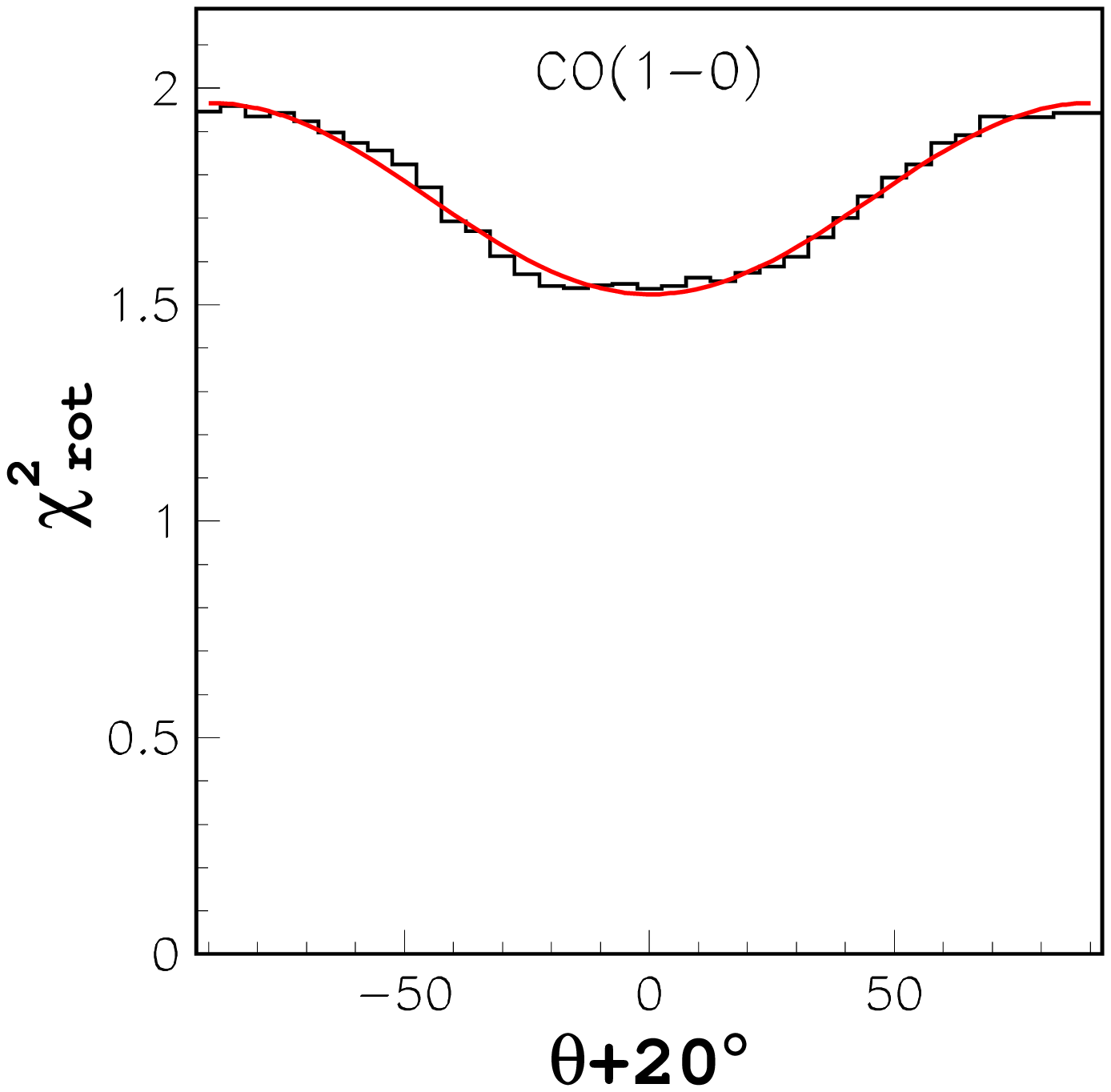}
\includegraphics[height=5cm,trim=0.5cm 0.7cm 0.cm 1.2cm,clip]{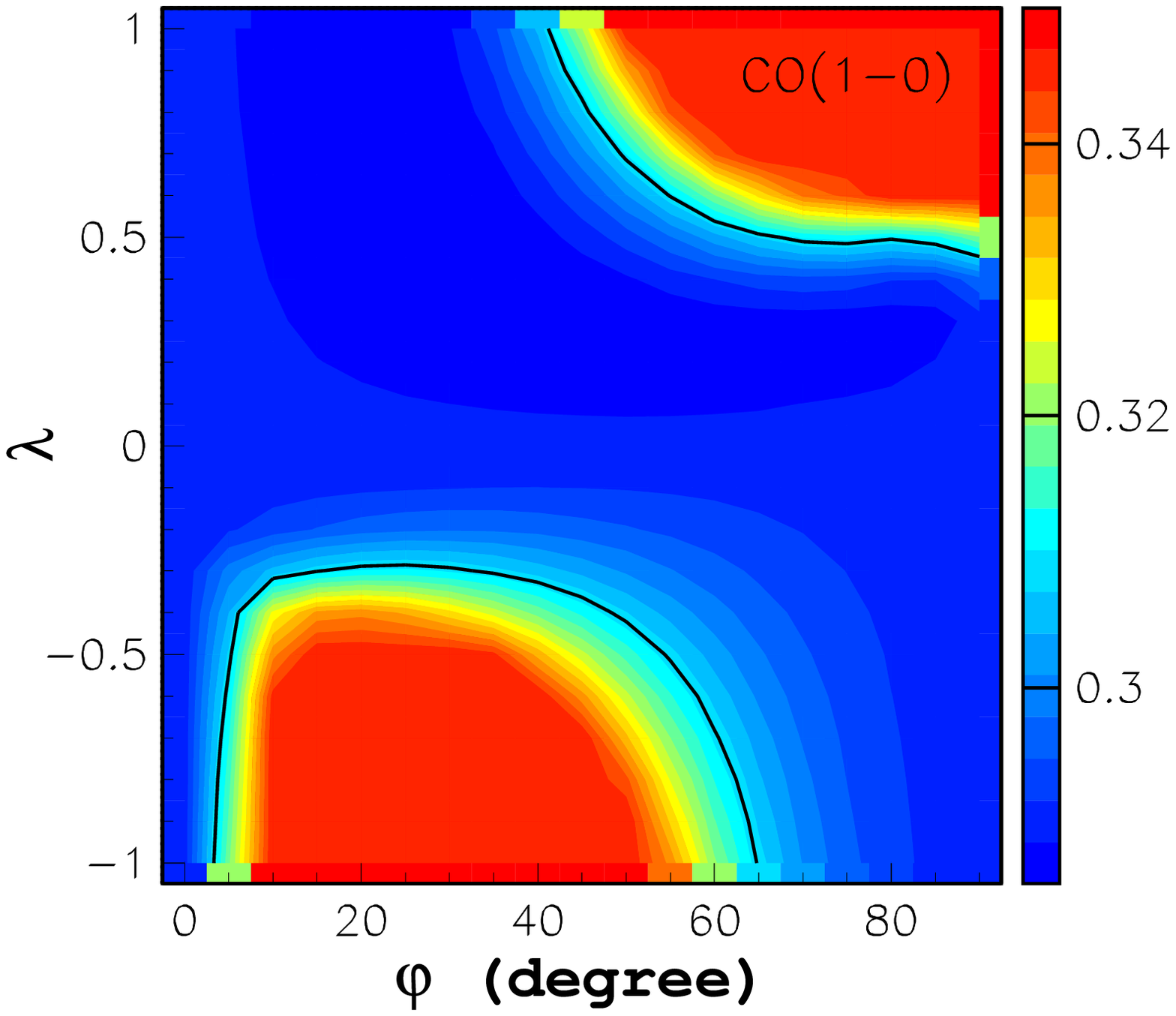}
\includegraphics[height=5cm,trim=0.5cm 0.7cm 0.cm 1.2cm,clip]{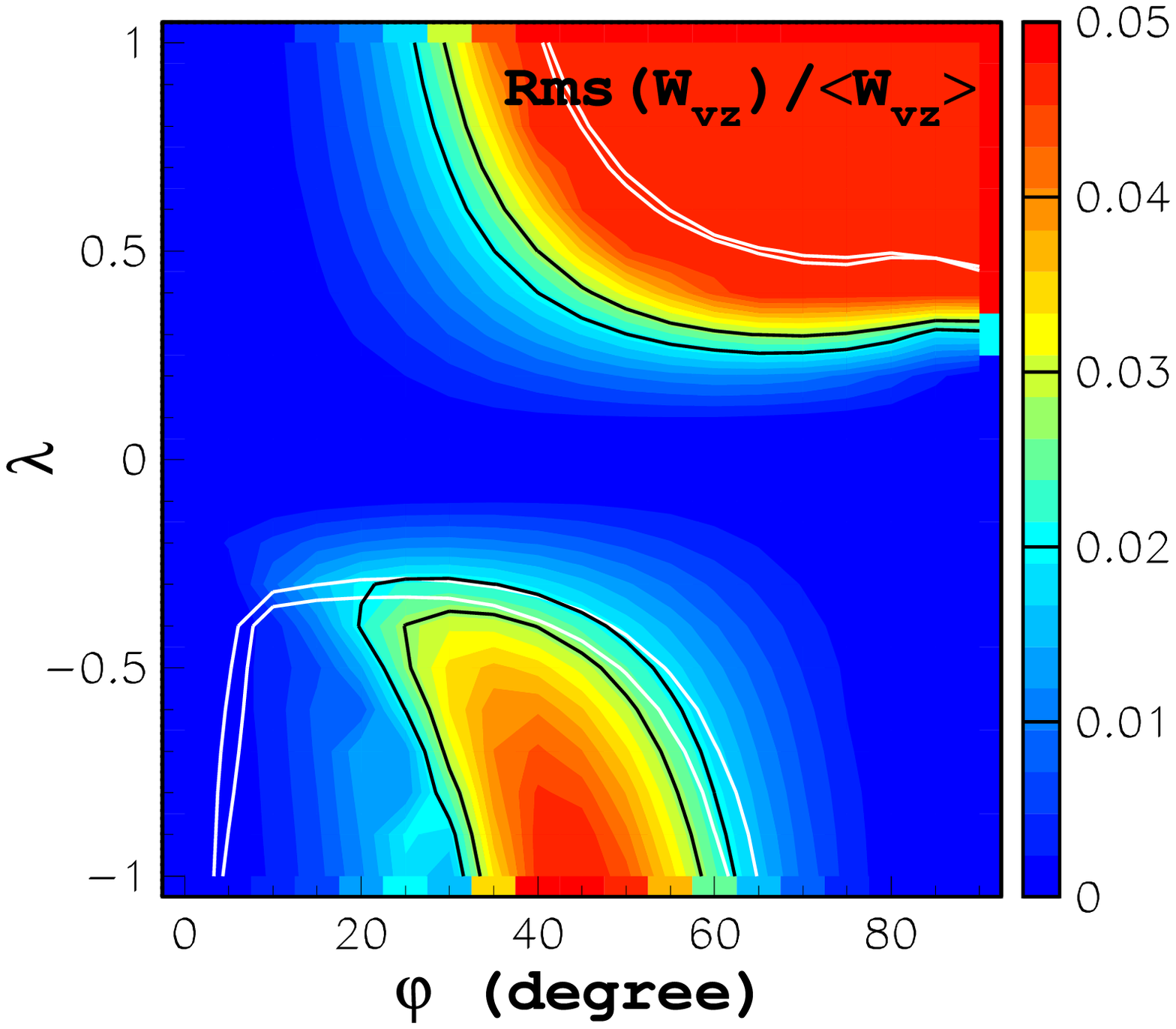}
\includegraphics[height=5cm,trim=0.5cm 0.7cm 0.9cm 1.2cm,clip]{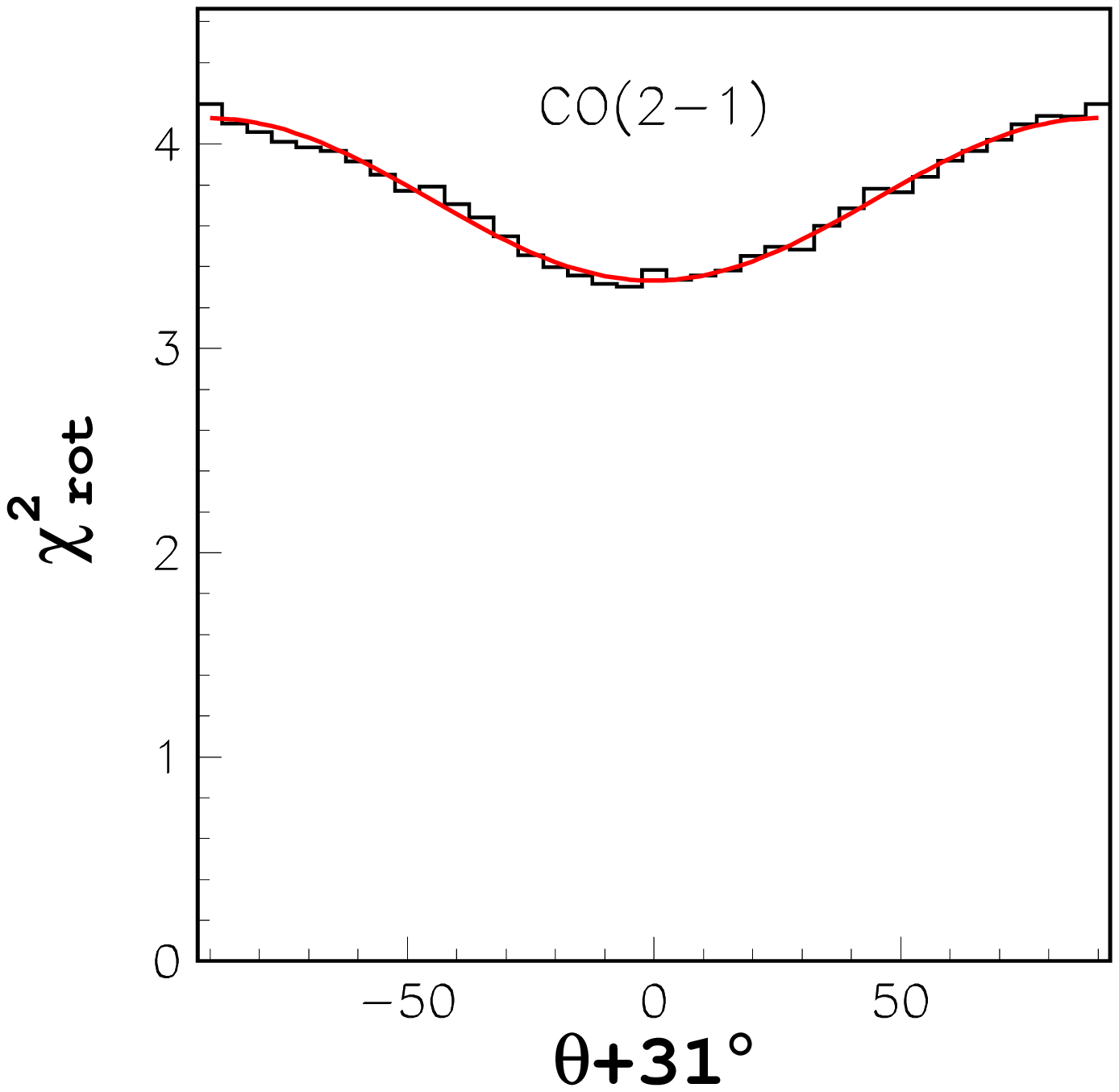}
\includegraphics[height=5cm,trim=0.5cm 0.7cm 0.cm 1.2cm,clip]{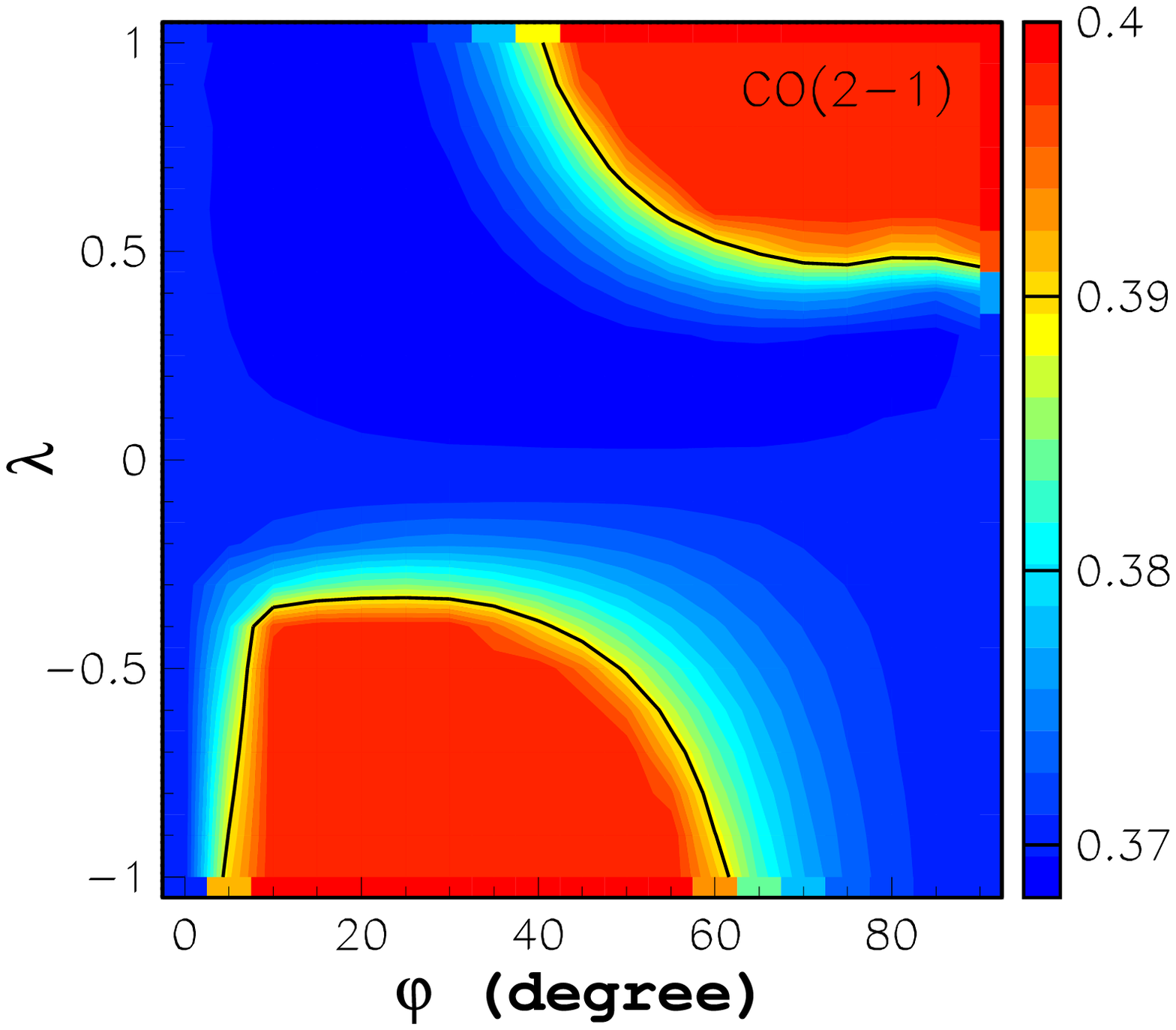}
\includegraphics[height=5cm,trim=0.5cm 0.7cm 0.cm 1.2cm,clip]{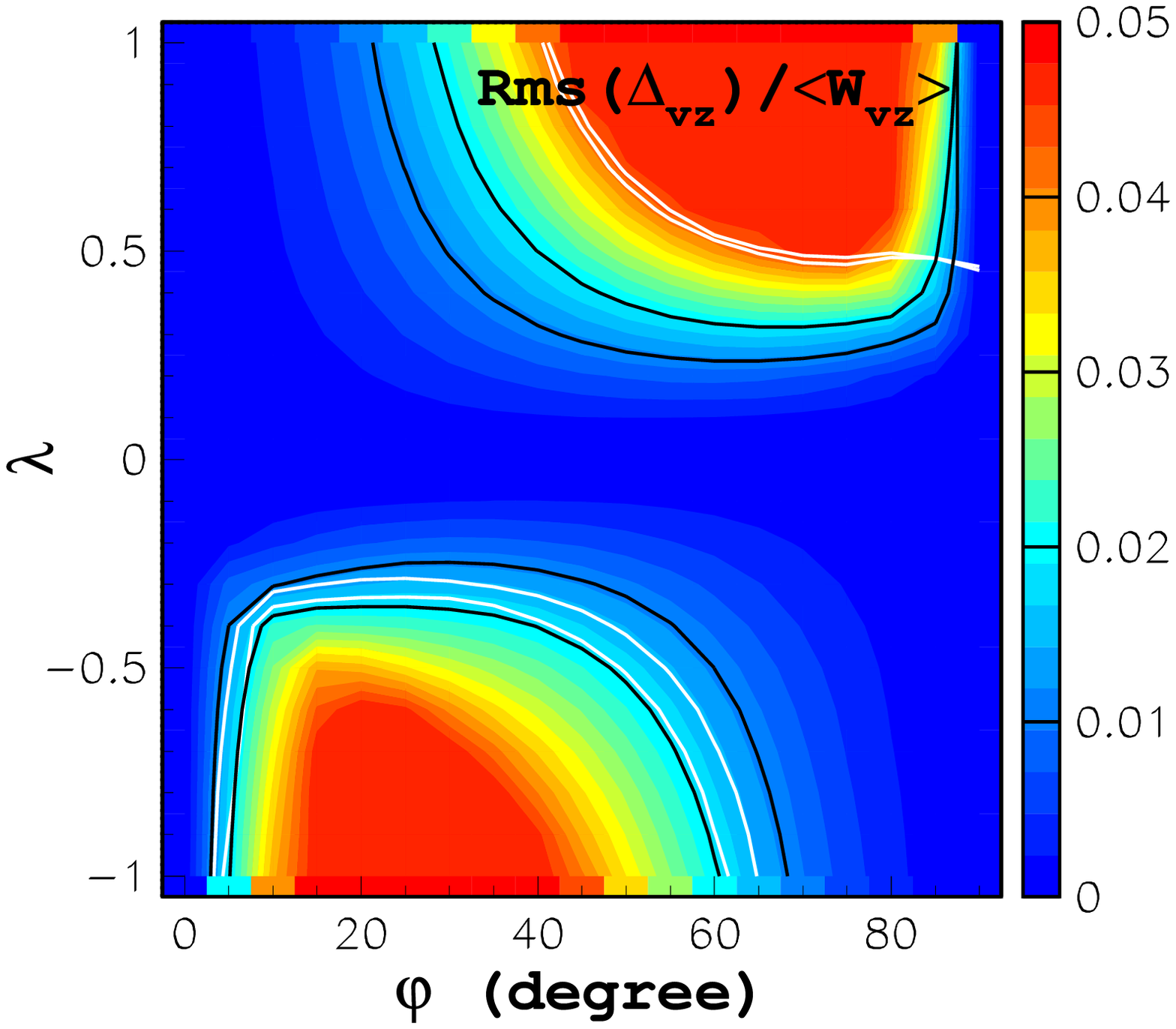}
\caption{EP Aqr CO(1-0) and CO(2-1) emission. Left: dependence of $\chi^2_{rot}$ on $\theta-\theta_0$; the curves are cosine square fits. Middle: ($\lambda,\varphi$) maps of $Q$ for the region $R<8$ arcsec. The contours show $Q=0.31$ for CO(1-0) and $Q=0.39$ for CO(2-1). Right: ($\lambda,\varphi$) maps of Rms($W_{vz}$)/$<W_{vz}>$ and Rms($\Delta_{vz}$)/$<W_{vz}>$ as predicted by the simple model. The black contours display the values measured for CO(1-0) and CO(2-1) emission. The white contours are the same as shown in black in the middle panels.}
\label{fig14}
\end{figure*}

We restrict the analysis to the sky plane region having $R<8$ arcsec. Figure \ref{fig13} displays sky maps of the deviation from unity of the ratio $\Delta(x,y)=F(x,y)/<F(x,y)>$ between the measured intensity $F(x,y)$ and its average $<F(x,y)>$ over position angle. It gives evidence for approximate isotropy in the sky plane, suggesting that the wind is either spherical or axi-symmetric with the symmetry axis near the line of sight. Indeed, $\chi^2_{rot}$ is found to display a very broad minimum, centered at $\theta_0 \sim  -20^\circ$ for CO(1-0) emission and at $\theta_0 \sim -31^\circ$ for CO(2-1) emission; its dependence on $\theta - \theta_0$ is shown in Figure \ref{fig14}. Using the same criterion as for RS Cnc to measure the uncertainty on its measurement, namely a 10\% increase of $\chi^2_{rot}$, gives $\theta= -20^\circ\pm 37^\circ$ and $-31^\circ \pm 43^\circ$ respectively, consistent with a common value of $-25^\circ\pm 28^\circ$. The agreement between the CO(1-0) and CO(2-1) results suggests that the deviation from sphericity is real and the broad distribution of $\chi^2_{rot}$ as a function of $\theta-\theta_0$ indicates that the star axis must make a small angle with the line of sight. Also shown in Figure \ref{fig14} (middle panels) is the dependence of $Q$ on $\lambda$ and $\varphi$. It remains nearly constant over a very large region of the ($\lambda,\varphi$) plane. The reason is that $V_{zmin}$ and $V_{zmax}$ are nearly invariant over the sky plane. In order to quantify better this statement, we integrate the Doppler velocity spectrum between $R=1$ arcsec and $R=8$ arcsec and display it in 18 bins of position angle $\psi$, each 20$^\circ$ wide. The high sensitivity that results allows for measuring in each bin $V_{zmin}$ and $V_{zmax}$ with a precision of $\sim 0.1$ \kms. We obtain this way measurements of the width $W_{vz}$ and the offset $\Delta_{vz}$ of the Doppler velocity spectrum in each bin of $\psi$. Their mean and rms values for CO(1-0) and CO(2-1) emission respectively, relative to the spectrum width, are
$<W_{vz}>=21.30$ \kms\ and 21.61 \kms, Rms($W_{vz}$)/$<W_{vz}>=2.1\ 10^{-2}$ and 2.9 10$^{-2}$ and Rms($\Delta_{vz}$)/$<W_{vz}>=1.2\ 10^{-2}$ and 2.0 10$^{-2}$ respectively. The latter two quantities are compared with the predictions of the simple model in the right panels of Figure \ref{fig14}. Compared with $Q$ they are more reliable discriminants in rejecting regions of the ($\lambda,\varphi$) plane but leave a very broad acceptable region along the descending diagonal. The situation here is very different from what was found for RS Cnc: it would not have helped to use a discriminant such as $P/T$ because the $V_0$ distribution is extremely narrow in the low $Q$ region of the ($\lambda,\varphi$) plane.
\begin{figure*}
\centering
\includegraphics[width=0.37\textwidth,trim=0.cm 0.cm 0.cm 0.cm,clip]{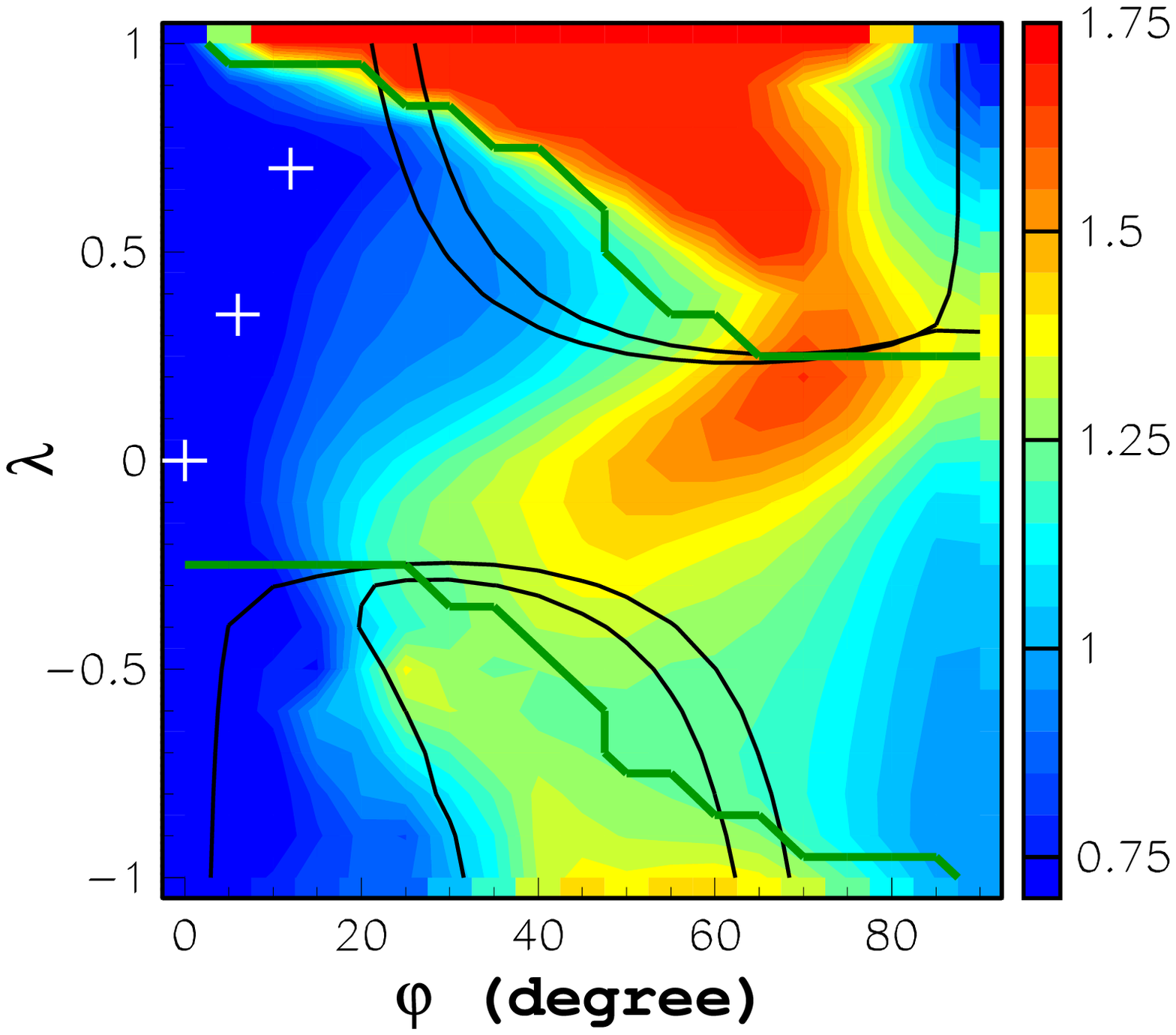}
\includegraphics[width=0.37\textwidth,trim=0.cm 0.cm 0.cm 0.cm,clip]{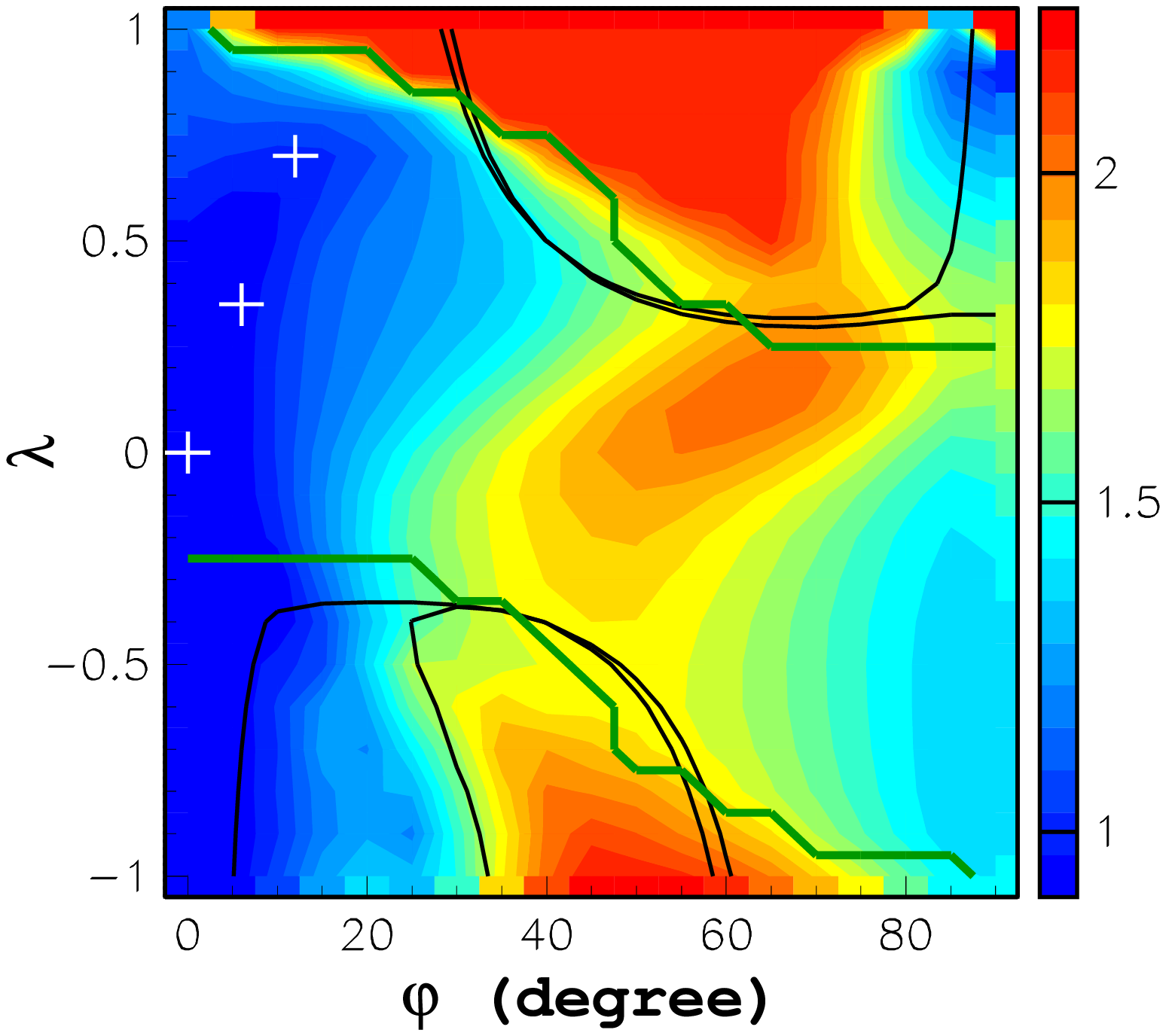}
\caption{EP Aqr CO(1-0) (left) and CO(2-1) (right) emission. Dependence of $\chi^2_{axi}$ on $\lambda$ and $\varphi$. On both panels the colour scale runs from minimum to 2.5 times minimum. The black contours indicate the regions rejected by Rms($W_{vz}$)/$<W_{vz}>$ and Rms($\Delta_{vz}$)/$<W_{vz}>$. The green contours show the limits of ambiguous de-projection. Crosses indicate values of the ($\lambda,\varphi$) pair listed in Table 2.}
\label{fig15}
\end{figure*}
Figure \ref{fig15} displays the dependence of $\chi^2_{axi}$ on $\lambda$ and $\varphi$. We use as uncertainty on the measured brightness the quadratic sum of 20\% of its value and of the noise. For both CO(1-0) and CO(2-1) emissions, large values of $\varphi$ are excluded; taking into account the constraint imposed by the study of $V_{zmin}$ and $V_{zmax}$ leaves a large region in the upper left quadrant of the map acceptable for de-projection.  The result obtained earlier by \cite{Nhung2015b}, ($V_0, \lambda, \theta, \varphi$)=(6.0 \kms, 0.67, $-36^\circ$, 13$^\circ$) is contained in this region. We select three representative ($\lambda,\varphi$) pairs for purpose of illustration: (0, 0$^\circ$), (0.35, 6$^\circ$) and (0.7, 12$^\circ$). The corresponding parameters are listed in Table 2 and the dependence on $r$, $\omega$ and $\sin\alpha$ of the associated de-projected effective emissivity multiplied by $r^2$ inside the sphere $r<8$ arcsec is displayed in Figure \ref{fig16}. The difference between the CO(1-0) and CO(2-1) radial distributions is understood as being caused by the different temperature dependence \citep[]{Nhung2018}. The smaller $\lambda$, the larger $V_0$ and the stronger the concentration of the effective emissivity near the equatorial plane. Indeed one evolves from a bipolar outflow for $\lambda = 0.7$ with no significant equatorial enhancement of effective emissivity to a spherical wind producing a strong enhancement of effective emissivity at equator for $\lambda = 0$. Only physics arguments can help choosing between the possible wind configurations. In particular, the presence of a companion star or massive planet in the equatorial plane would probably favour an enhancement of the effective emissivity while not making the wind velocity much deviate from spherical. A bipolar wind associated with an isotropic effective emissivity would be more difficult to explain. However, such arguments are well beyond the scope of the present article.

\begin{figure*}
\centering
\includegraphics[width=0.9\textwidth,trim=0.cm 0.cm 0.cm 0.cm,clip]{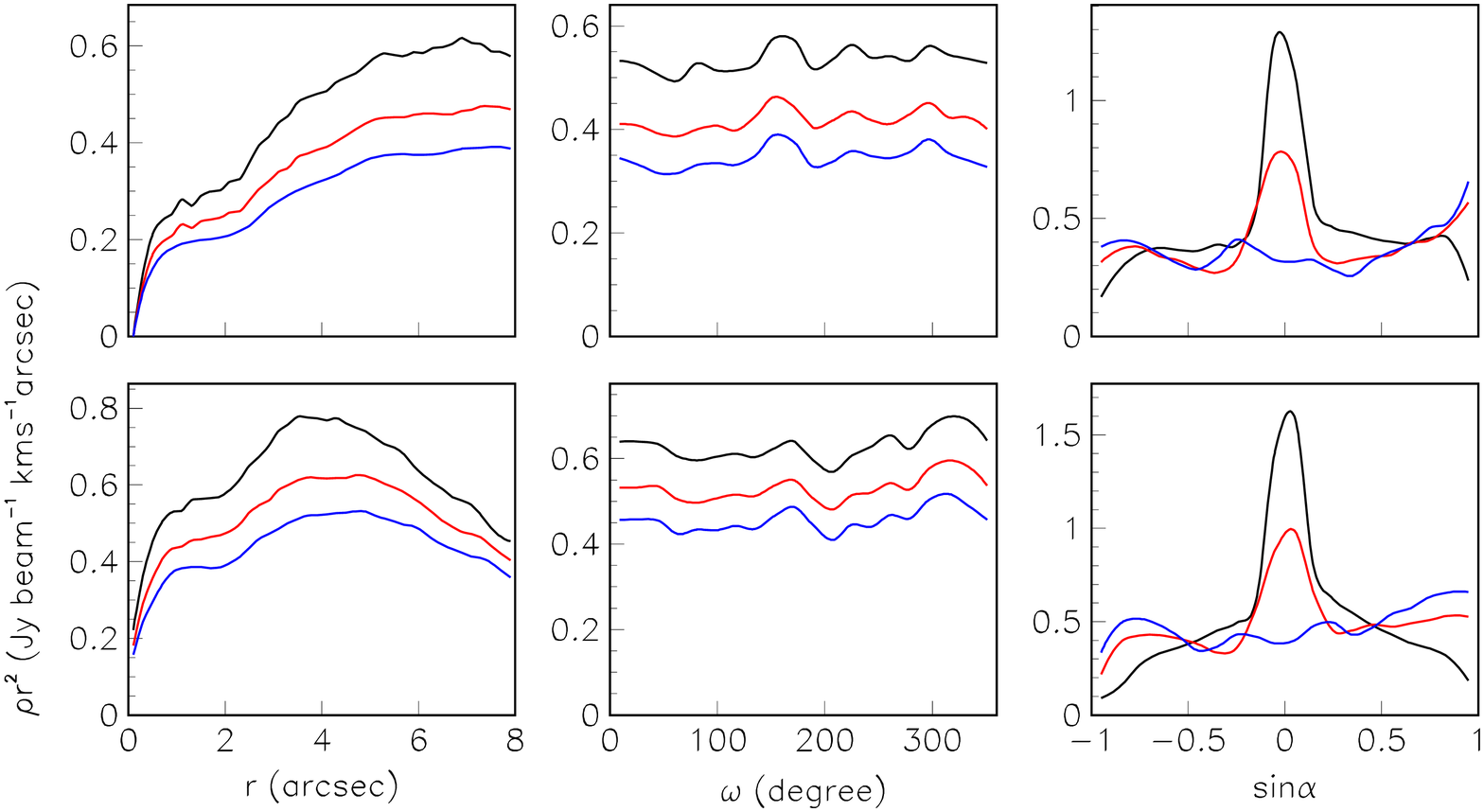}
\caption{Dependence of the de-projected emissivity for EP Aqr multiplied by $r^2$ (Jy beam$^{-1}$\kms arcsec) on $r$ (arcsec, left), $\omega$ (degrees, middle) and $\sin \alpha$ (right) respectively. The values of ($\lambda, \varphi$) are, from up down, (0, 0$^\circ$), (0.35, 6$^\circ$) and (0.7, 12$^\circ$), corresponding to the crosses in Figure \ref{fig15}. Upper panels are for CO(1-0) emission and lower panels for CO(2-1) emission.}
\label{fig16}
\end{figure*}
\begin{table*}
  \centering  
  \caption{De-projection of EP Aqr CO(1-0) observations: summary of results}    
  \label{tab2}        
  \renewcommand{\arraystretch}{1.1}
  \begin{tabular}{|c|c|c|c|c|c|c|c|}
    \hline
    Emission  &	Case  &	$V_0 $(\kms)  &	$\lambda$    &	$\varphi$ ($^\circ$)   &	$\theta$ ($^\circ$)    &	$Q$          &$\chi^2_{axi}$\\
    
    \hline
    CO(1-0)\&(2-1) & \citet{Nhung2015b}&6.0&	0.67 &  	13         &  $-36$                   &	$-$           &	$-$	\\
    \hline
    \multirow{3}{*}{CO(1-0)}&1         &10.6     &0             &0    &\multirow{3}{*}{$-$20}    &\multirow{3}{*}{0.29}&0.44 \\
    \cline{2-5}\cline{8-8}
                             &2        &8.0      &0.35          &6    &                                &              &0.57 \\
    \cline{2-5}\cline{8-8}
                            &3         &6.5      &0.7           &12   &                                &              &0.66 \\
    \hline
    \multirow{3}{*}{CO(2-1)}&1         &10.8     &0              &0   &\multirow{3}{*}{$-$31}     &\multirow{3}{*}{0.37}&0.68 \\
    \cline{2-5}\cline{8-8}
                            &2        &8.1       &0.35          &6       &                            &                &0.89 \\
    \cline{2-5}\cline{8-8}
                             &3          &6.6        &0.7        &12     &                           &                  &1.00 \\
    
    \hline
        
  \end{tabular}
\end{table*}

\section{SUMMARY AND CONCLUSION}
The above analysis of the problem of de-projection of radially expanding axi-symmetric circumstellar envelopes has provided a deep insight into its main features and has devised tools that help with its solution. However, it does not offer a substitute to a detailed analysis that takes into account the physics of the mechanisms at play, such as hydrodynamical constraints and radiative transfer considerations. It is only meant to shed light on some of its intricacies and to set the frame for preliminary considerations that can help the construction of a realistic physical model. The effective emissivity used throughout the paper to describe observations is a convenient quantity that offers simplicity but hides the difficulty of disentangling the effects of temperature from those of density. Its assumed axi-symmetry is often a good approximation but is only valid for envelopes that are sufficiently optically thin and the assumption of radial expansion, implying the absence of rotation, is often violated in the later stages of the star evolution on their way to planetary nebulae. It is therefore essential to keep in mind the limited scope of the results of the present work.\\

With these caveats in mind, we summarize below the main results that have been obtained:\\

i) The position angle $\theta$ of the projection of the star axis on the sky plane has been shown to minimize a quantity, $\chi^2_{rot}$, independently from the particular form taken by the effective emissivity; its minimization, as long as it is made over the whole data-cube and not simply on its projection on the sky plane, makes optimal use of the available information. Only in cases where the problem has no solution, either because the star axis is close to the line of sight or because the wind velocity and the effective emissivity are nearly isotropic, is $\theta$ ill-defined.   \\

ii) A good evaluation of the scale of the wind velocity $-V_0$ in the simple model$-$has been obtained from a study, in each pixel, of the dependence on position angle $\psi$ of the end points $V_{zmin}$ and $V_{zmax}$ of the Doppler velocity spectra. To this effect, the rms deviation of the velocity scale relative to its mean over the sky plane, $Q$, was found to be sufficient, in the case of simulated data, to quantify the agreement with a model. Its minimization helps in eliminating regions of the prolateness ($\lambda$) versus inclination ($\varphi$) plane which are unsuitable for de-projection. The method is particularly efficient when the wind velocity is confined near the sky plane, either as strongly prolate with axis close the sky plane or as strongly oblate with axis close to the line of sight. However, in the case of real data, a more careful analysis is needed in order to exploit the richness of the information contained in the dependence of $V_{zmin}$ and $V_{zmax}$ on $\lambda$ and $\varphi$. A careful evaluation of $V_{zmin}$ and $V_{zmax}$, taking the noise level in proper account, is mandatory and algorithms allowing for separating the $V_0$ peak from a background of improper values may be helpful.  

iii) Having obtained sensible estimates of $\theta$ and $V_0$, one can further constrain the ($\lambda,\varphi$) pair by imposing axi-symmetry on the de-projected effective emissivity. To this effect, a quantity $\chi^2_{axi}$ has been constructed, which is minimal for maximal axi-symmetry.
Contrary to the minimization of $Q$, the minimization of $\chi^2_{axi}$ is inefficient when the wind velocity is confined near the sky plane. In such cases, the evaluation of the optimal value of the ($\lambda,\varphi$) pair relies more on the constraints imposed by the distribution on the sky plane of the width and offset of the Doppler velocity spectra than on the constraints imposed by the requirement of axi-symmetry.

iv) Having obtained estimates of the orientation of the star axis, $\theta$ and $\varphi$, of the scale of the wind velocity, $V_0$, and of the effective prolateness of its distribution, $\lambda$, we are then in a position to de-project the effective emissivity as long as a single value of $z/r$ is associated with each bin of measured Doppler velocity. However, the effective emissivity cannot be directly de-projected in regions of $z/r$ that are associated with ambiguous velocity bins. Such regions are particularly important when the wind velocity is confined in the vicinity of the sky plane, implying that the Doppler velocity distribution is folded on itself and confined to lower values. In such cases, the observations measure mostly the projection of the effective emissivity on the sky plane, namely the integrated brightness (or intensity). 

v) To a good approximation, the $r$-dependence of the de-projected effective emissivity is obtained independently from the choice of the wind configuration used for de-projection as long as the wind velocity does not depend on $r$. \\

vi) As an illustration, we have presented two case studies of CO emission of AGB stars. They were chosen to be representative of typical observations rather than of the best space and spectral resolutions and sensitivity available today. In both cases, the method and tools developed in the present article have been shown to select efficiently wind configurations suitable for de-projection. Published analyses of the relevant observations have proposed models that are indeed favoured by these results. However, they also suggest exploring other wind configurations, of different inclination and prolateness, which may deserve being considered in the framework of a physics analysis. More importantly, they provide a deep understanding of the constraints imposed on a physics model and of how unique is the region of the ($\lambda,\varphi$) plane ultimately selected.

vii) The main contribution of the present work may be the insight it has provided on the issue of the under-determination of the problem of de-projection. It has underlined the importance of being conscious that a broad family of wind configurations, illustrated in blue in the right panel of Figure \ref{fig3}, can be used in principle to de-project the observed brightness data-cube into an effective emissivity at each point in space. In practice, however, constraints imposed by the requirement of axi-symmetry of the de-projected effective emissivity and of the need to populate in each pixel the totality of the observed Doppler velocity spectrum (but no more) have been found to complement each other; taken together, they are efficient in restricting the domain of the ($\lambda,\varphi$) plane acceptable for de-projection. A good approximation to the $r$-distribution of the de-projected emissivity has been obtained under the hypothesis of constant velocity but physics considerations must then be used to decide how best to combine a possible velocity gradient with such $r$-dependence. More generally, the constraints imposed by hydrodynamics on the relation between density, temperature and velocity will be determinant in deciding on the ``best'' physics model.

\section*{ACKNOWLEDGEMENTS}
We thank Professor Pierre Lesaffre for a careful reading of the manuscript and very useful and pertinent comments that helped improving significantly its content. 
This paper makes use of the following ALMA data: 2016.1.00026.S. ALMA is a partnership of ESO (representing its member states), NSF (USA) and NINS (Japan), together with NRC (Canada), NSC and ASIAA (Taiwan), and KASI (Republic of Korea), in cooperation with the Republic of Chile. The Joint ALMA Observatory is operated by ESO, AUI/NRAO and NAOJ. It also makes use of observations carried out with the IRAM NOEMA Interferometer and the IRAM 30 m telescope. IRAM is supported by INSU/CNRS (France), MPG (Germany) and IGN (Spain). This research is funded by Graduate University of Science and Technology under grant number GUST.STS.DT2017-VL01.
Financial and/or material support from the Vietnam National Space Center, the National Foundation for Science and Technology Development (NAFOSTED), the World Laboratory and Odon Vallet fellowships is gratefully acknowledged.\\

%%%%%%%%%%%%%%%%%%%%%%%%%%%%%%%%%%%%%%%%%%%%%%%%%%



%\bibliographystyle{mnras}


%%%%%%%%%%%%%%%%%%%%%%%%%%%%%%%%%%%%%%%%%%%%%%%%%%

%%%%%%%%%%%%%%%%% APPENDICES %%%%%%%%%%%%%%%%%%%%%

%%%%%%%%%%%%%%%%%%%%%%%%%%%%%%%%%%%%%%%%%%%%%%%%%%

\appendix

% Don't change these lines
\bsp	% typesetting comment
\label{lastpage}
\end{document}